\documentclass[graybox]{svmult}

\usepackage{mathptmx}       
\usepackage{helvet}         
\usepackage{courier}        
\usepackage{type1cm}        
\usepackage{makeidx}         
\usepackage{graphicx}        
\usepackage{multicol}        
\usepackage[bottom]{footmisc}

\usepackage{natbib}
\usepackage{cite}

\newcommand{\Msun}{{\ensuremath{\mathrm{M}_{\odot}}}}
\newcommand{\Zsun}{{\ensuremath{\mathrm{Z}_{\odot}}}}
\newcommand{\cp}{\ensuremath{\mathrm{\xi}_{2.5}}}

\makeindex             

\begin{document}

\title*{The Deaths of Very Massive Stars}
\author{S. E. Woosley and Alexander Heger}
\institute{Stan Woosley \at Department of Astronomy and Astrophysics,
  UCSC, Santa Cruz CA 95064 USA \\ \email{woosley@ucolick.org} \and
  Alexander Heger \at School of Mathematical Sciences, Monash
  University, Victoria 3800,
  Australia \\ \email{alexander.heger@monash.edu}}
%
%
\maketitle

\vskip -1.3 in

\abstract{The theory underlying the evolution and death of stars
  heavier than 10 \Msun \ on the main sequence is reviewed with an
  emphasis upon stars much heavier than 30 \Msun. These are stars
  that, in the absence of substantial mass loss, are expected to
  either produce black holes when they die, or, for helium cores
  heavier than about 35 \Msun, encounter the pair instability. A wide
  variety of outcomes is possible depending upon the initial
  composition of the star, its rotation rate, and the physics used to
  model its evolution. These heavier stars can produce some of the brightest
  supernovae in the universe, but also some of the faintest. They can
  make gamma-ray bursts or collapse without a whimper. Their
  nucleosynthesis can range from just CNO to a broad range of elements
  up to the iron group. Though rare nowadays, they probably played a
  disproportionate role in shaping the evolution of the universe
  following the formation of its first stars.}

\section{Introduction}
\label{sec:intro}

Despite their scarcity, massive stars illuminate the universe
disproportionately. They light up regions of star formation and stir
the media from which they are born. They are the fountains of element
creation that make life possible. The neutron stars and black holes
that they make are characterized by extreme physical conditions that
can never be attained on the earth. They are thus unique
laboratories for nuclear physics, magnetohydrodynamics, particle
physics, and general relativity. And they are never quite so
fascinating as when they die.

Here we briefly review some of aspects of massive star death. The
outcomes can be crudely associated with three parameters - the star's
mass, metallicity, and rotation rate. In the simplest case of no
rotation and no mass loss, one can delineate five outcomes and assign
approximate mass ranges (in some cases {\sl very} approximate mass
ranges) for each. These masses then become the section heads for the
first part of this chapter.  1) From 8 to 30 \Msun \ on the main
sequence (presupernova helium core masses up to 12 \Msun), stars
mostly produce iron cores that collapse to neutron stars leading to
explosions that make most of today's observable supernovae and heavy
elements. Within this range there are probably islands of stars that
either do not explode or explode incompletely and make black holes,
especially for helium cores from 7 to 10 \Msun. 2) From 30 to 80 \Msun
\ (helium core mass 10 to 35 \Msun), black hole formation is quite
likely. Except for their winds, stars in this mass range may be
nucleosynthetically barren. Again though there will be exceptions,
especially when the effects of rotation during core collapse are
included. 3) 80 to (very approximately) 150 \Msun \ (helium cores 35
to 63 \Msun), pulsational-pair instability supernovae. Violent
nuclear-powered pulsations eject the star's envelope and, in some
cases, part of the helium core, but no heavy elements are ejected and
a massive black hole of about 40 \Msun \ is left behind. 4) 150 - 260
\Msun \ (again very approximate for the main sequence mass range, but
helium core 63 to 133 \Msun), pair instability supernovae of
increasing violence and heavy element synthesis. No gravitationally
bound remnant is left behind. 5) Over 260 \Msun \ (133 \Msun \ of
helium), with few exceptions, a black hole consumes the whole star.
Rotation generally shifts the main sequence mass ranges (but not the
helium core masses) downwards for each outcome. Mass loss complicates
the relation between initial main sequence mass and final helium core
mass.

The latter part of the paper deals with some possible effects of rapid
rotation on the outcome. In the most extreme cases, gamma-ray bursts
are produced, but even milder rotation can have a major affect on the
light curve and hydrodynamics if a magnetar is formed.

\section{The Deaths of Stars 8 \Msun \ to 80 \Msun}

\subsection{Compactness as a Guide to Outcome}
\label{sec:compact}

The physical basis for distinguishing stars that become supernovae
rather than planetary nebulae, and that are therefore, in some sense,
``massive'', is the degeneracy of the carbon-oxygen (CO) core
following helium core burning. Stars with dense, degenerate CO cores
develop thin helium shells and eject their envelopes leaving behind
stable white dwarfs, while heavier stars go on to burn carbon and
heavier fuels. A mass around 8 \Msun \ is usually adopted for the
transition point. The effects of degeneracy linger, however, on up to
at least 30 \Msun \ at oxygen ignition, and to still heavier masses for
silicon burning. Even at 80 \Msun, the center of a massive star has
become degenerate by silicon depletion.

Were the core fully degenerate and composed of nuclei with equal
numbers of neutrons and protons, its maximum mass would be the cold
Chandrasekhar mass, 1.38 \Msun. This cold Chandrasekhar mass is
altered however, both by electron capture reactions, which tend to
reduce it, and the high temperatures necessary to burn oxygen and
silicon, which increase it \citep{Cha39,Hoy60,Tim96}. For main sequence
stars from 8 to 80 \Msun, the iron core mass at the time it collapses
varies from about 1.3 to 2.3 \Msun \ (baryonic mass), with the larger
values appropriate for more massive stars. Surrounding this degenerate
core is a nested structure of shells that cause adjustments to the
density structure. For very degenerate cores with energetic shells at
their edges, the presupernova structure resembles that of an
asymptotic giant branch star - a compact core surrounded by thin
burning shells and a low density envelope with little gravitational
binding energy. The matter outside of the iron core is easily ejected
in such stars, and it is easy to make a supernova out of them, even
with an inefficient energy source like neutrinos. Heavier stars with
less degenerate cores and shells farther out, on the other hand, have
a density that declines more slowly. These mantles of heavy elements,
where ultimately most of the nucleosynthesis occurs, are more tightly
bound and the star is more difficult to blow up.

\begin{figure}[h]
\includegraphics[angle=90,scale=0.45]{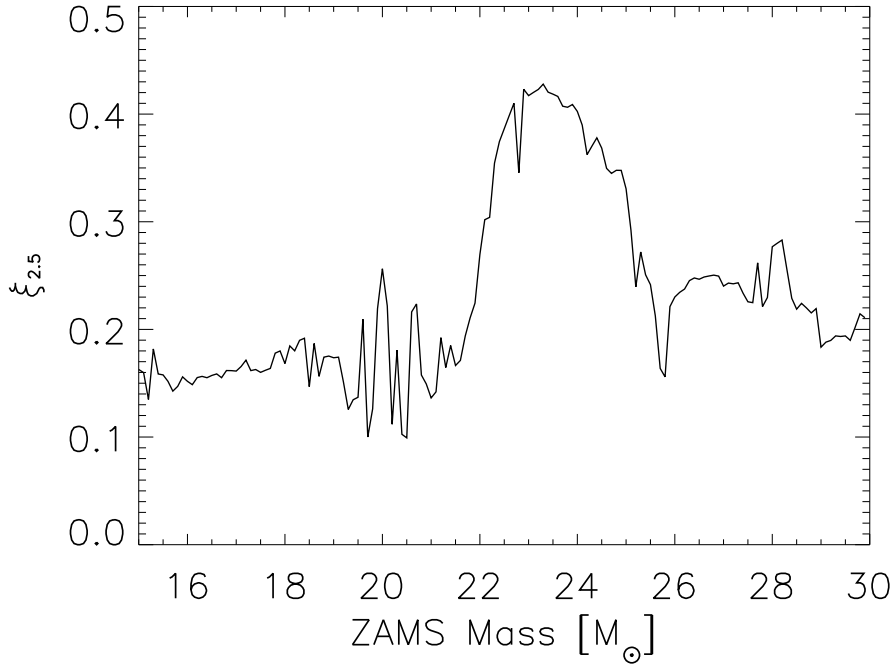} 
\caption{Compactness parameter for presupernova stars of solar
  metallicity as a function of main sequence mass \citep{Suk14}. Stars
  with smaller $\zeta_{2.3}$ explode more easily.}
\label{fig:compact}      
\end{figure}

\citet{Oco11} have defined a ``compactness parameter'', \cp =
2.5/R$_{2.5}$, that is a quantitative measure of this density fall
off. Here R$_{2.5}$ is the radius, in units of 1000 km, of the mass
shell in the presupernova star that encloses 2.5 \Msun. The fiducial
mass is taken to be well outside the iron core but deep enough in to
sample the density structure around that core. It makes little
difference whether this compactness is evaluated at the onset of
hydrodynamical instability or at core bounce \citep{Suk14}.  Figure
\ref{fig:compact} shows \cp \ as a function of main sequence mass for
stars of solar metallicity. O'Connor and Ott and \citet{Ugl12} have
both shown that it becomes difficult to explode the star by neutrino
transport alone if \cp \ becomes very large. The critical value is not
certain and may vary with other properties of the star, but in
Ugliano's study is usually 0.20 to 0.30. By this criterion, it may be
difficult to explode stars in the 22 to 24 \Msun \ range (at least) as
well as all stars above about 30 \Msun \ that do not lose substantial
mass along the way to their deaths. The latter especially includes
stars with very subsolar metallicity.

There are a number of caveats that go along with this speculation.
The structure of a presupernova star is not fully represented by a
single number and its compactness is sensitive to a lot of stellar
physics, including the treatment of semiconvection and convective
overshoot mixing and mass loss and the nuclear reaction rates employed
\citep{Suk14}. Rotation and magnetic fields will change both the
presupernova structure and its prospects for explosion by non-neutrino
processes. Finally, the surveys of how neutrino-powered explosions
depend on compactness have, so far, been overly simple and mostly in
1D, though see recent work by Janka and colleagues
\citep{Jan12a,Jan12b,Mul12}.  Still the simplification introduced by
this parametrization is impressive and reasonably consistent with
what we know about the systematics of supernova progenitors.

%

Figure \ref{fig:compact} suggests that stars below 22 \Msun \ should
be, for the most part, easy to explode using neutrinos alone and no
rotation. This is consistent with the observational limits that
\citet{Sma09} and \citet{Sma09b} placed upon about a dozen
presupernova progenitor masses as well as the estimated mass of SN
1987A. It also is a minimal set of masses if the solar abundances are
to be produced \citep{Bro13}.  The compactness of stars between 22 and
about 35 \Msun \ is highly variable though due to the migration
outwards of the carbon and oxygen burning shells \citep{Suk14}. For a
standard choice of stellar physics, there exists an island of compact
cores between 26 and 30 \Msun \ that might allow for islands of
``explodability''. This would help with nucleosynthesis and also
possibly have implications for the properties of the Cas A supernova
remnant.  Cas A, like SN 1993J and 2001gd \citep{Che10}, is thought to
be the remnant of a relatively massive single star that lost most of
its hydrogenic envelope either to a wind or a binary companion, yet
its remnant contains a neutron star. If the mass loss was to a
companion, as is currently thought, then the progenitor mass was
probably less than 20 \Msun, but if a star of 30 \Msun \ could explode
after losing most of its envelope, this might provide an alternate,
solitary star explanation.

On the other hand, binary x-ray sources exist and the black holes in
them are thought to be quite massive \citep{Oze10,Wik14}. Stars above 35
\Msun \ either make black holes if their mass loss during the
Wolf-Rayet stage is small, or some variant of Type Ibc supernovae if
it is large and shrinks the carbon oxygen core below about 6 \Msun.

Probably the greatest omission here is the effect of rotation and the
need to produce gamma-ray bursts in a subset of stars. We also have
said nothing about the fate of stars over 80 \Msun. Both topics will be
covered in later sections.


\subsection{8 \Msun \ to 30 \Msun; Today's Supernovae and Element Factories}
\label{sec:10to80}

For reasonable choices of initial mass function, stars in this mass
range are responsible for most of the supernovae we see today and for
the synthesis of most of the heavy elements. This does not preclude
many of these stars from making black holes, but the supernovae we see
are in this range. Baring binary interaction, including mergers, or
low metallicity, such stars are, at death, red supergiants, and so the
most common supernovae are Type IIp. Explosion energies range from 0.5
to $4 \times 10^{51}$ erg with a typical value of $9 \times 10^{50}$
erg \citep{Kas09}.  These values are the kinetic energy of all ejecta
at infinity and the actual energy requirement for the central engine
may be larger, especially for more massive stars with large binding
energies in their mantles. The light curves and spectra of the models
are consistent with observations, to the extent that models for SN IIp
can even be used as ``standard candles'' based upon the expanding
photosphere method.

Including binary interactions, one can account for the remainder of
common (non-thermonuclear) supernovae, including Type Ib, Ic, IIb, etc
\citep{Des11,Des12}. These events typically come from massive stars in
the 12 - 18 \Msun \ range that lose their binary envelopes and die as
stripped down helium cores of 3 to 4 \Msun. On the low end, the
explosion ejects too little $^{56}$Ni to be a bright optical
event. Heavier stars are rarer and may not explode. If they do their
light curves are broader and fainter than typical Ib and Ic
supernovae. 

The nucleosynthesis produced by solar metallicity stars in this mass
range has been explored many times
\citep{Woo95,Woo02,Woo07b,Thi96,Nom06,Lim00,Chi04,Chi13,Hir05,Nom13}.
While the results from the different groups studying the problem
vary depending upon the treatment of critical reaction rates, mass
loss, semiconvection, convective overshoot, and rotationally induced
mixing, some general conclusions may be noted.

\begin{itemize}

\item The majority of the elements and their isotopes from carbon (Z =
  6) through strontium (Z = 38) are made in solar proportions in
  supernovae with an average production factor of around 15 (IMF
  averaged yield expressed as a mass fraction and divided by the
  corresponding solar mass fraction). The iron group, Ti through Ni,
  is underproduced in massive stars by a factor of several, which is
  consistent with the premise that most of the solar abundances of
  these species were made recently in thermonuclear (Type Ia)
  supernovae. In the distant past, the oxygen to iron ratio was
  larger, and massive stars probably produced the iron group in very
  low metallicity stars.

\item For a reasonable choice for the critical
  $^{22}$Ne($\alpha,$n)$^{25}$Mg reaction rate, the light s-process up
  to A = 90 is made well in massive stars, but only if the upper bound
  for the masses of stars that explode is not too low
  \citep{Bro13}. The heavy component of the p-process above A = 130 is
  also produced in massive stars, but the production of the lighter
  p-process isotopes (A = 90 - 130) remains a mystery, especially the
  origin of the abundant closed shell nucleus $^{92}$Mo (Z = 42, N =
  50).

\item While oxygen is definitely a massive star product, the elemental
  yield of carbon ($^{12}$C) is sensitive to how mass loss is treated
  and requires for its production the inclusion of the winds of stars
  heavier than 30 \Msun. Red giant winds, AGB mass loss, and planetary
  nebulae also produce $^{12}$C, perhaps most of it, as well as all of
  $^{13}$C and $^{14}$N. $^{15}$N and $^{17}$O are not sufficiently
  produced in massive stars and may be made in classical novae.

\item $^{11}$B and about one-third of $^{19}$F are made by neutrino
  spallation in massive star supernovae. $^6$Li, $^9$Be, and $^{10}$B
  do not appear to be substantially made, and probably owe their
  origin to cosmic ray spallation in the interstellar medium. Some but
  not all of $7$Li is made by neutrino spallation.

\item Certain select nuclei like $^{44}$Ca, $^{48}$Ca, and $^{64}$Zn are
  underproduced and may require alternate synthesis

\end{itemize}

In addition to the previously mentioned uncertainties affecting
presupernova evolution, assumptions about the explosion mechanism also
play a major role. Fundamentally important is just which masses of
stars eject their mantles of heavy elements and which collapse to
black holes while ejecting little new elements. For a given
presupernova structure, a shock that imparts $\sim$10$^{51}$ erg of
kinetic energy to the base of the ejecta, none of which fall back,
will give a robust pattern of nucleosynthesis whether that energy is
imparted by a piston or as a thermal ``bomb''. The approximation used
by many, however, that the explosion across all masses can be
parametrized by a constant kinetic energy at infinity is too crude
and needs revisiting. Stars of different masses have different binding
energies, compactness parameters, and iron core masses. Rotation
probably has a major effect on the explosion, especially of the
more massive stars. The next stage of modeling will need to take into
account these dependencies.

\subsection{Stars 30 \Msun \ to 80 \Msun; Black Hole Progenitors}

While the jury is still out regarding the mass-dependent efficiency of
an explosion mechanism that includes realistic neutrino transport,
rotation, magnetic fields, and relativity in three dimensions, the
existence of stellar mass black holes and the absence of observable
supernova progenitors with high mass implies that at least some stars
do not explode and eject all of their heavy element inventory. Until
such time as credible models exist, a reasonable assumption is that
the success of the explosion is correlated with the compactness
\citep{Oco11,Ugl12}. By this criterion, one expects the central regions
of stars with {\sl helium cores} much larger than about 10 \Msun \ and
lighter than 35 \Msun \ to collapse \citep{Suk14} to black holes. Above
10 \Msun \ of helium, or about 30 \Msun \ on the main sequence, the
iron core is large, typically over 2.0 \Msun \ and the compactness
parameter is large. Above 35 \Msun, or about 80 \Msun \ on the main
sequence, one encounters the pulsational pair instability (Section
~\ref{sec:80to150}).

For solar metallicity stars, mass loss may reduce the presupernova
mass of the star to a level where it can frequently explode. If it
does and the entire envelope has been lost, the explosion will be some
sort of Type Ib or IC supernova. Because of the large mass, the light
curve would be broad, and not as bright as most observed SN Ibc. The
remnant would probably be a neutron star. It is unclear if such events
have been observed, though Cas A might be a candidate.

Even if the core of the star collapses to a black hole, its death is
not necessarily nucleosynthetically barren or unobservable. The black
hole could result from fall back and the envelope may still be
ejected. Even if the presupernova star does not explode at all, its
evolution will still have contributed to nucleosynthesis by its wind,
which may be appreciable \citep{Hir05}. If only the hydrogenic layers
are ejected, these winds can be a rich source of $^{12}$C, $^{16}$O
and, at low metallicity, $^{14}$N \citep{Mey02}. If the wind eats
deeply into the helium core, $^{18}$O and $^{22}$Ne can also be
ejected, but the winds of such stars are devoid of heavier elements
like silicon and iron.

If the star rotates sufficiently rapidly, a gamma-ray burst may result
(Section \ref{sec:grb}) or a magnetar-powered supernova. Even for
non-rotating stars, it is debatable whether the star can simply
disappear without a trace.  The sudden loss of mass energy from the
protoneutron star can trigger mass ejection and a very subluminous
supernova \citep{Lov13}.  Pulsations or gravity waves generated in the
final stages of evolution may partly eject the envelope. Even a weak
explosion might produce a potentially observable bright spike as its
shock wave erupts through the surface of the star \citep{Pir13}. In a
tidally locked binary or a low metallicity blue supergiant with
diminished mass loss, sufficient angular momentum may exist in the
outermost layers of the star to pile up in an accretion disk around
the new black hole producing some sort of x-ray and gamma-ray
transient \citep{Woo12,Qua12b}.

\subsection{Yesterday's Metal Poor Stars}
\label{sec:lowz} 

Stars with lower metallicity, as may have predominated in the early
universe, can have different presupernova structures for a variety of
reasons \citep{Suk14}. Most importantly, metallicity affects mass loss,
especially for the more massive stars. If the amount of mass lost is
low or zero, the presupernova star including its helium core, is
larger, and that has a dramatic effect on its compactness and
explodability. A vastly different outcome is expected for e.g., a 60
\Msun \ star that retains most of its hydrogen envelope and dies with
a helium core of 24 \Msun, and one that loses all of its envelope as
well as most of its helium core to die with a total mass of 7.3
\Msun. This small mass is obtained with current estimates of mass loss
for solar metallicity stars \citep{Woo02}. Indirect effects can also
come into play.  Because a low metallicity star loses less mass, it
loses less angular momentum and thus dies rotating more
rapidly. Indeed, there is some suggestion from theory that massive
stars are all born rotating near break up and only slow as a
consequence of evolution (expansion) and mass loss \citep{Ros12}.

Very low metallicity may also enhance the probability of forming more
massive stars \citep{Abe02}. Whether this results in much more massive
stars than are being born today is being debated. While this is an
important issue for the frequency of first generation stars with
masses over 80 \Msun \ (Section~\ref{sec:80to150}), an equally
important question is whether the IMF for the first generation stars
might have been ``bottom-light'', that is producing a deficiency of
stars below some characteristic mass, say $\sim$30 \Msun
\citep{Tan04}. Since this would remove the range of masses responsible
for most supernovae and nucleosynthesis today, the early universe
would have been quite a different place.

Even assuming the exact same masses of stars and explosions as today,
nucleosynthesis would be distinctly different in low metallicity
stars.  The amount of neutrons available to produce all isotopes
except those with Z = N depends on the ``neutron excess'', $\eta =
\Sigma (N_i - Z_i)(X_i/A_i)$, where $Z_i$, $N_i$, and $A_i$ are the
proton number, neutron number and atomic weight of the species ``i''
and $X_I$ is its mass fraction. At the end of hydrogen burning all CNO
(essentially the metallicity of the star) has become $^{14}$N. Early
in helium burning this becomes $^{18}$O by the reaction sequence
$^{14}$N($\alpha,\gamma)^{18}$F($e^+ \nu)^{18}$O. The weak interaction
here is critical as it creates a net neutron excess that persists
throughout the rest of the star's life and limits the production of
neutron rich isotopes (like $^{22}$Ne, $^{26}$Mg, $^{30}$Si etc) and
odd-Z elements (like Na, Al, P). Other weak interactions in later
stages of evolution also increase $\eta$, so that by the time one
reaches calcium, the dependence on initial metallicity is not so
great, but one does expect an affect on the isotopes from oxygen
through phosphorus.

Assuming that the IMF was unchanged and using the same explosion model
as for solar metallicity stars (but suppressing mass loss) gives an
abundance set that agrees quite well with observations of
metal-deficient stars in the range -4 $<$ [Z/\Zsun] $<$ -2
\citep{Lai08}. All elements from C through Zn are well fit without the
need for a non-standard IMF or unusually high explosion energy.

Below [Z/\Zsun] = -4, one becomes increasing sensitive to individual
stellar events and to the properties of the first generation stars.
If the stars below 30 \Msun \ are removed from the sample, the
nucleosynthesis is set by a) the pre-collapse winds of stars in the 30
- 80 \Msun \ range; b) the results of rotationally powered explosions
with uncertain characteristics; and c) the contribution of pulsational
pair- and pair-instability supernovae (see below). If only a) and c)
contribute appreciably, the resulting nucleosynthesis could be CNO
rich and very iron poor.

The light curves of metal deficient supernovae below 80 \Msun \ are
likely to be different - some of the time. If the stars die as red
supergiants, then very similar Type IIp supernovae will result,
but more of the stars are expected to die as blue supergiants with
light curves like SN 1987A \citep{Heg10}. Rotation can alter this
conclusion, however, as it tends to increase the number of red
supergiants compared with blue \citep{Mae12}. To the extent that the
massive stars retain their hydrogenic envelope, Type Ib and Ic
supernovae will be suppressed, though of course a binary channel
remains a possibility.

\section{Pulsational Pair Instability Supernovae (80 to 150 \Msun)}
\label{sec:80to150}

The pair instability occurs during the advanced stages of massive
stellar evolution when sufficiently high temperature and low density
lead to a thermal concentration of electron-positron pairs sufficient
to have a significant effect on the equation of state. Only the most
massive stars have sufficiently high entropy to encounter this
instability. Making the rest mass of the pairs in a post-carbon
burning star takes energy that might have otherwise contributed
to the pressure. As a result, for a time, the pressure does not rise
rapidly enough in a contracting stellar core to keep pace with
gravity. The structural adiabatic index of the core dips below 4/3
and, depending on the strength of the instability, the core contracts
more or less rapidly to higher temperature, developing considerable
momentum as it does so. As temperature rises, carbon, oxygen and, in
some cases, silicon burn rapidly. The extra energy from this burning,
plus the eventual partial recovery from the instability when the pairs
become highly relativistic, causes the pressure to rebound fast enough
to slow the collapse. If enough burning occurs before the infall
momentum becomes too great, the collapse is reversed and an explosion
is possible. For stars that are too big though, specifically for
helium non-rotating cores above 133 \Msun, the collapse continues to a
black hole.

When an explosion happens, it can be of two varieties. If enough
burning occurs to unbind the star in a single pulse, a
``pair-instability supernova'' results (Section~\ref{sec:pairsn}). If
not, the core of the star expands violently for a time and may kick
off its outer layers, including any residual hydrogen envelope. It
then slowly contracts until the instability is encountered again and
the core pulses once more. The process continues until enough mass has
been ejected and entropy lost as neutrinos that the pair instability
is finally avoided and the remaining star evolves smoothly to iron
core collapse. Typically this requires a reduction of the helium and
heavy element core mass to below 40 \Msun. These repeated
thermonuclear outbursts can have energies ranging from ``mild'',
barely able to eject even the loosely bound hydrogen envelope of a red
supergiant, to extremely large, with over 10$^{51}$ erg in a single
pulse. On the high energy end, collisions of ejected shells can
produce very bright transients. The observational counterpart is
``pulsational pair-instability supernovae'' (PPSN).

Depending upon rotation, the electron-positron pair instability begins
to have a marked effect on the post-carbon burning evolution of
massive stars with negligible mass loss when their main sequence mass
exceeds about 70 - 80 \Msun. (Extremely efficient
rotationally-induced mixing leading to chemically homogeneous chemical
evolution can reduce the threshold main sequence mass still further to
approximately the threshold helium core mass \citep{Cha12}). For solar
metallicity, stars this massive are usually assumed to lose all their
hydrogen envelope and part of their cores along the way and thus avoid
the instability. Suffice it to say that if the combined effects of
mass loss and rotation allow the existence of a helium core mass in
excess of 34 \Msun \ at carbon depletion, the pair instability will
have an effect. To get a full-up pair instability supernova, one needs
a helium core mass of about 63 \Msun \ which might correspond,
depending upon the treatment of convection physics, to a main sequence
star around 150 \Msun. In between, lies the PPSN. As we shall see, the
final evolution of such stars can be quite complicated because of the
many pulses, but they have the merit that the explosion hydrodynamics
is simple.

\subsection{Pulsationally Unstable Helium Stars}
\label{sec:heliumppsn}

\begin{figure}
\centering
\begin{tabular}{cc}
\includegraphics[scale=.22]{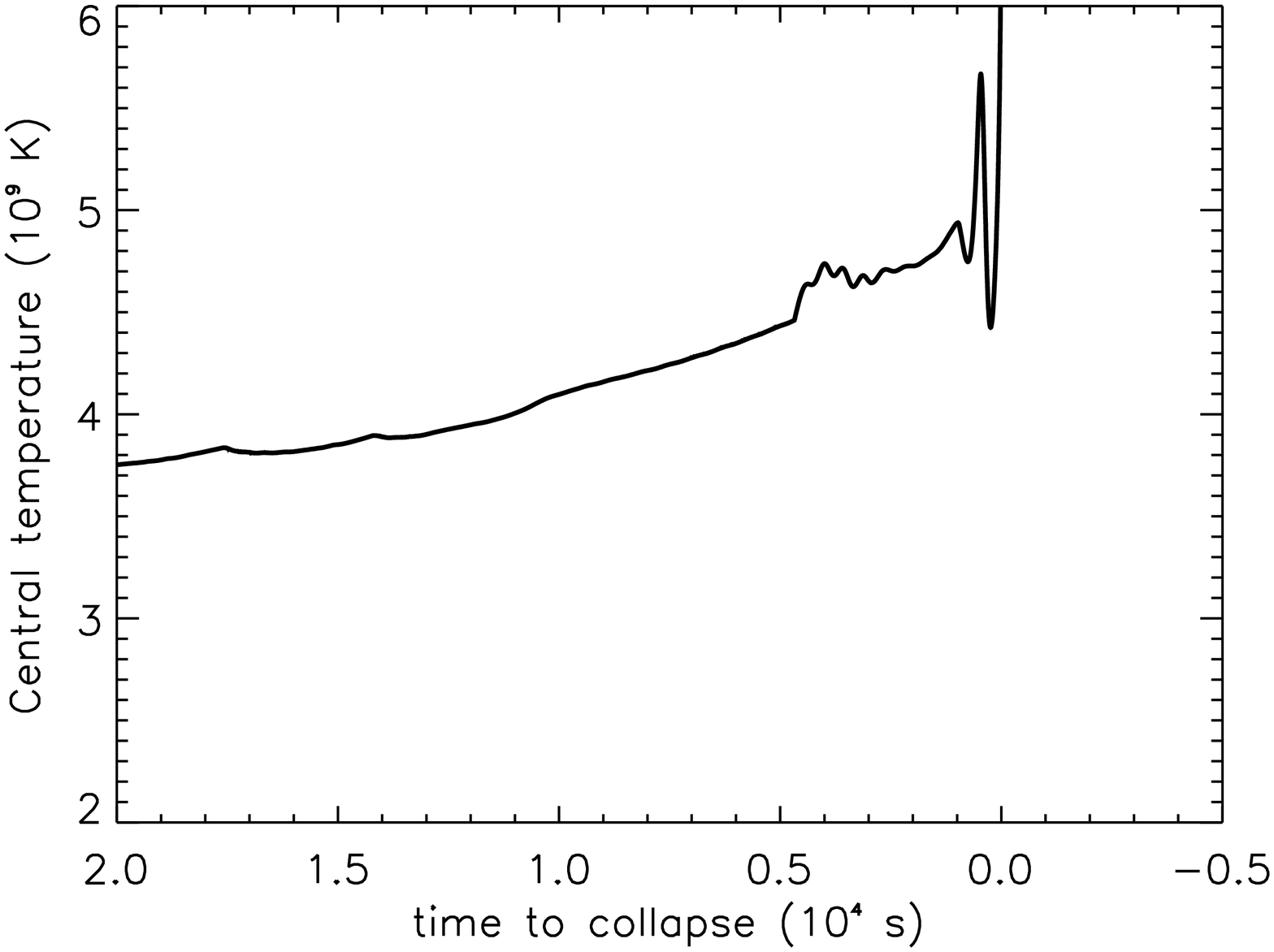} &
\includegraphics[scale=.22]{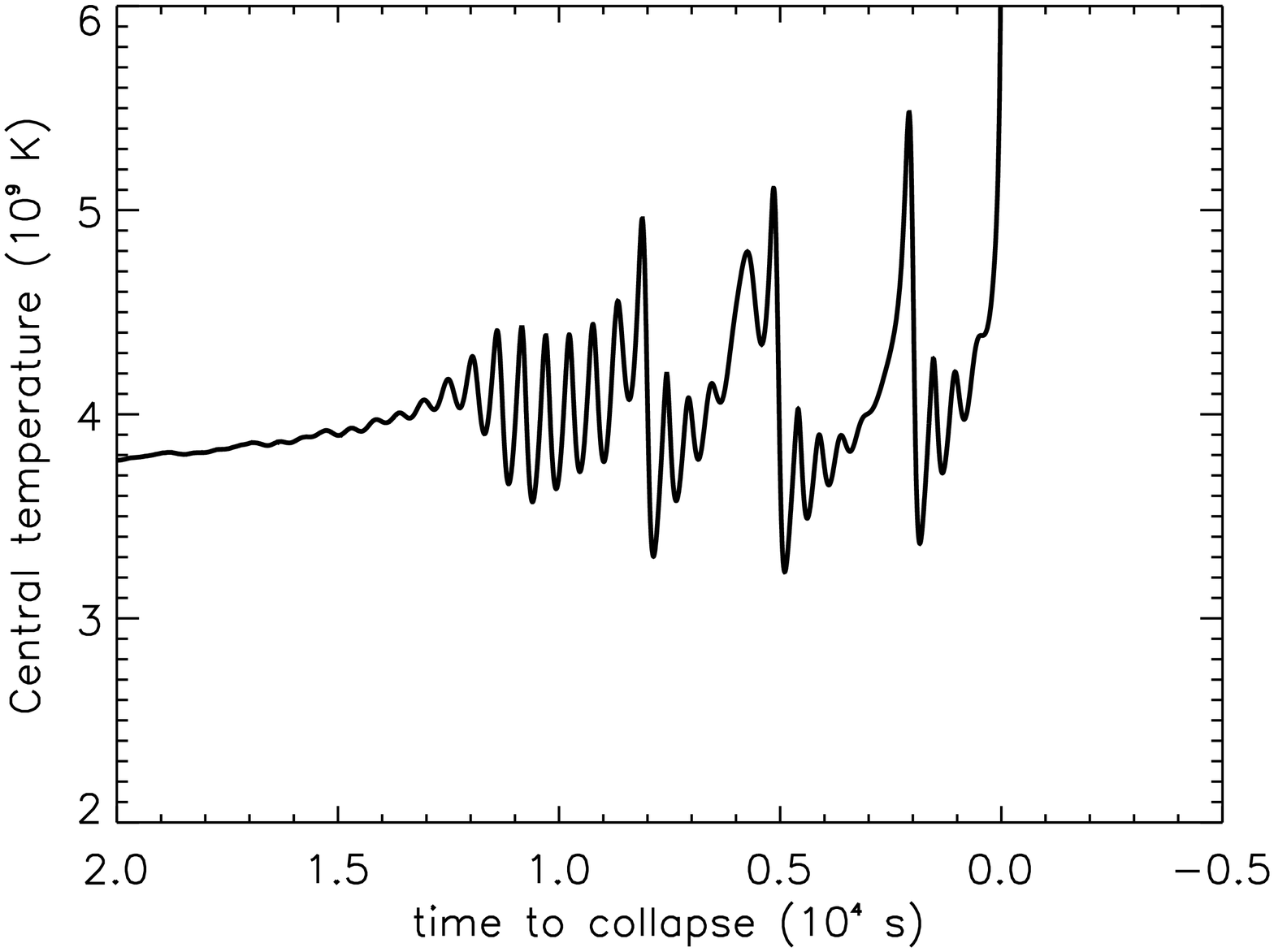} \\
\includegraphics[scale=.22]{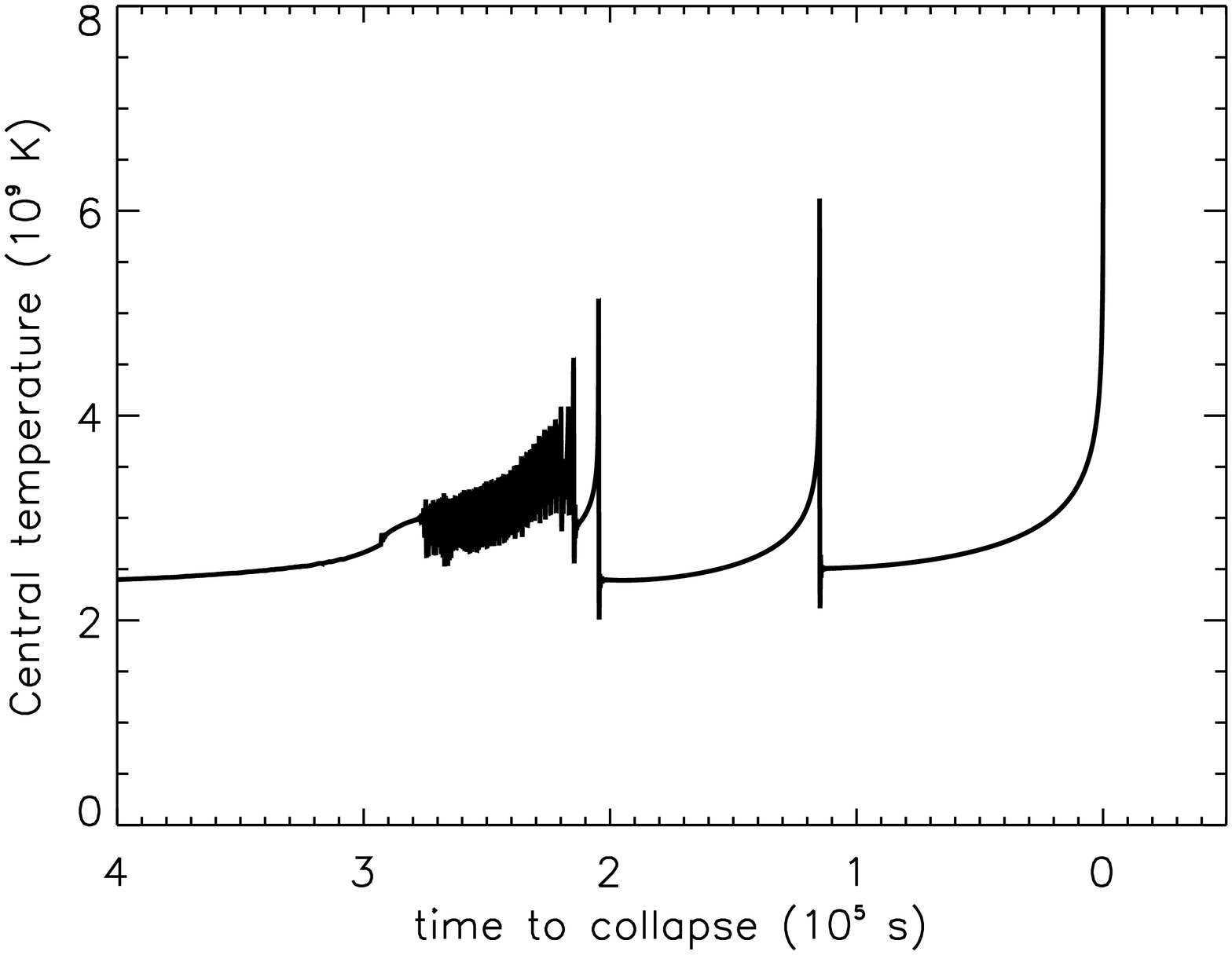} &
\includegraphics[scale=.22]{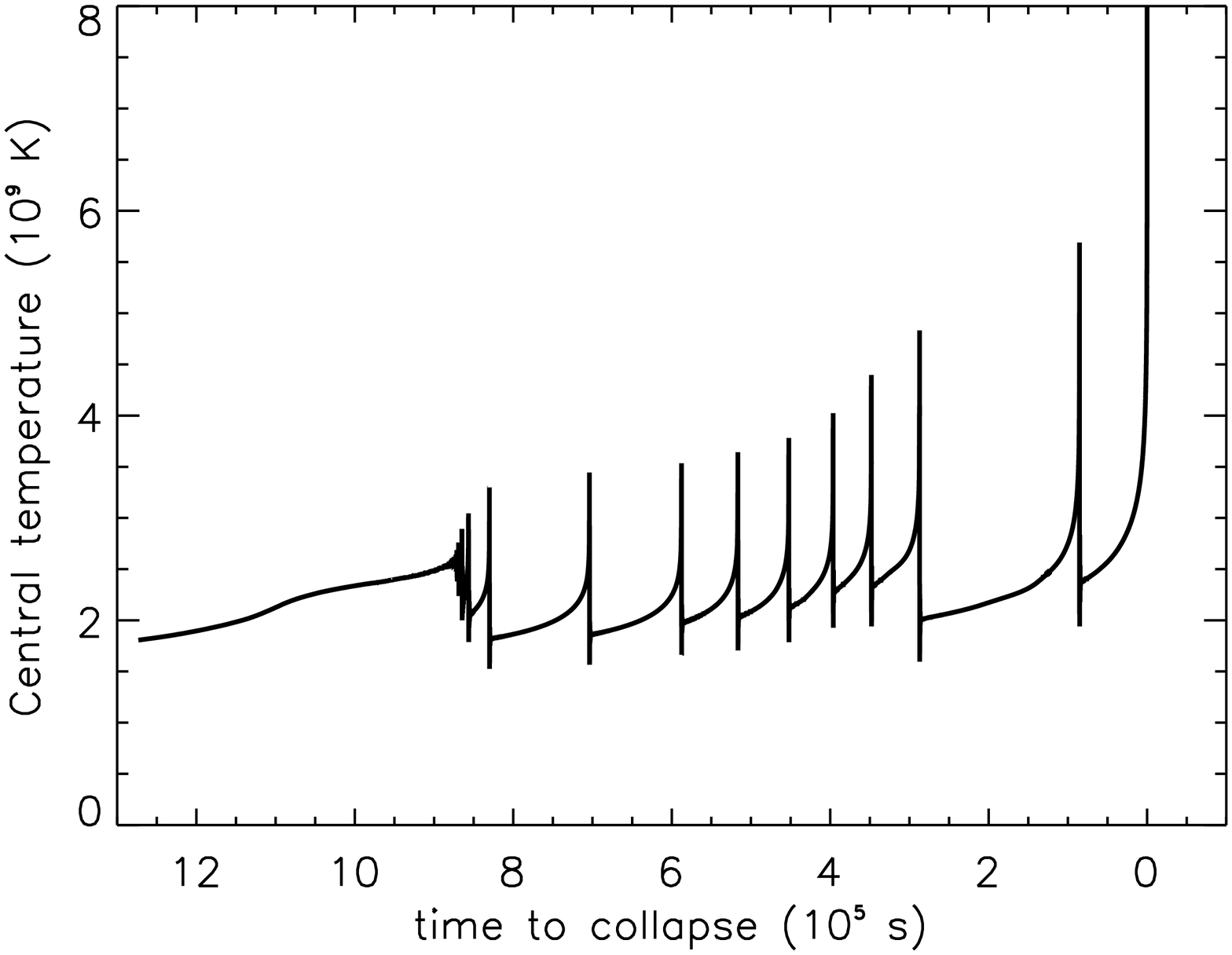} \\
\includegraphics[scale=.22]{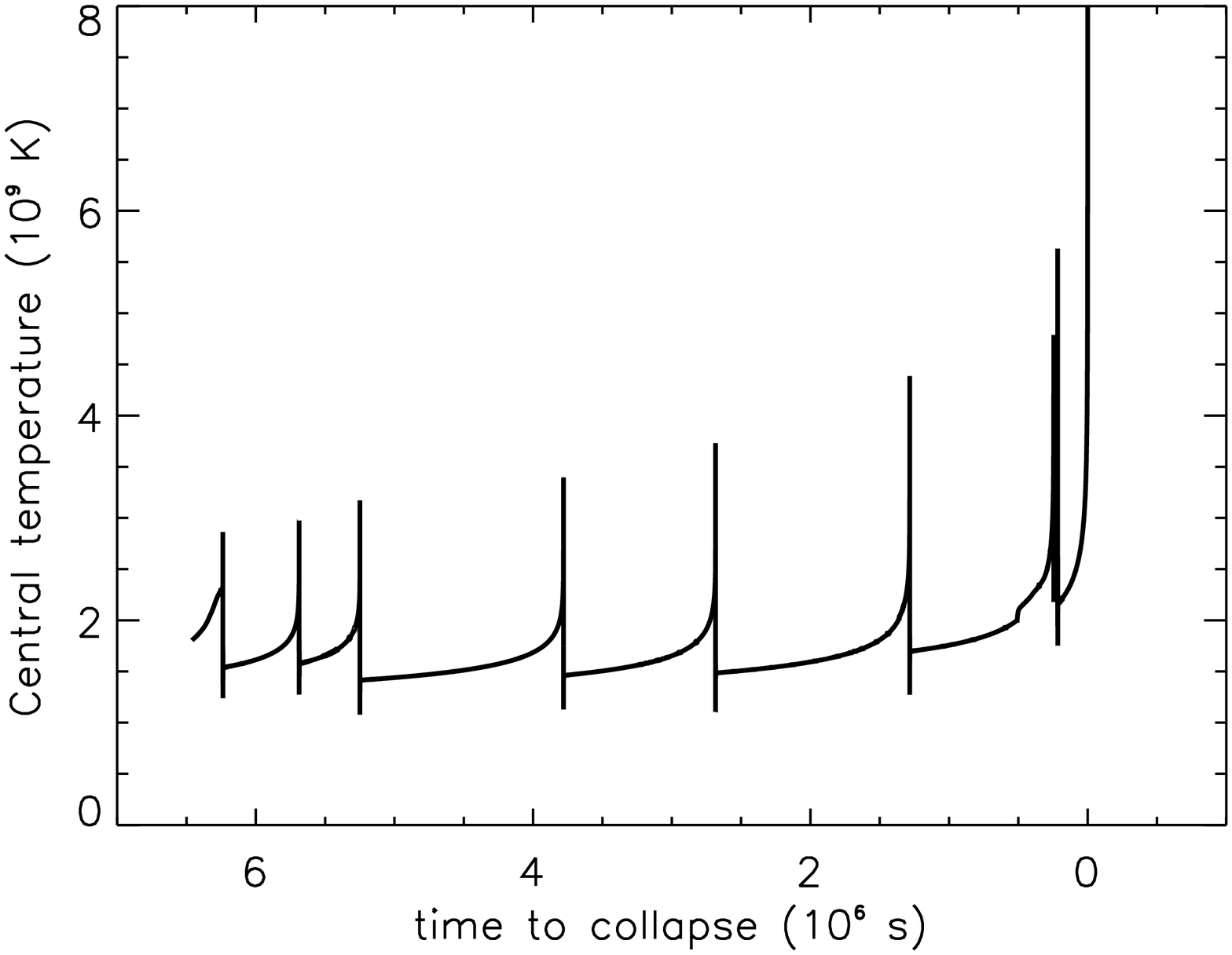} &
\includegraphics[scale=.22]{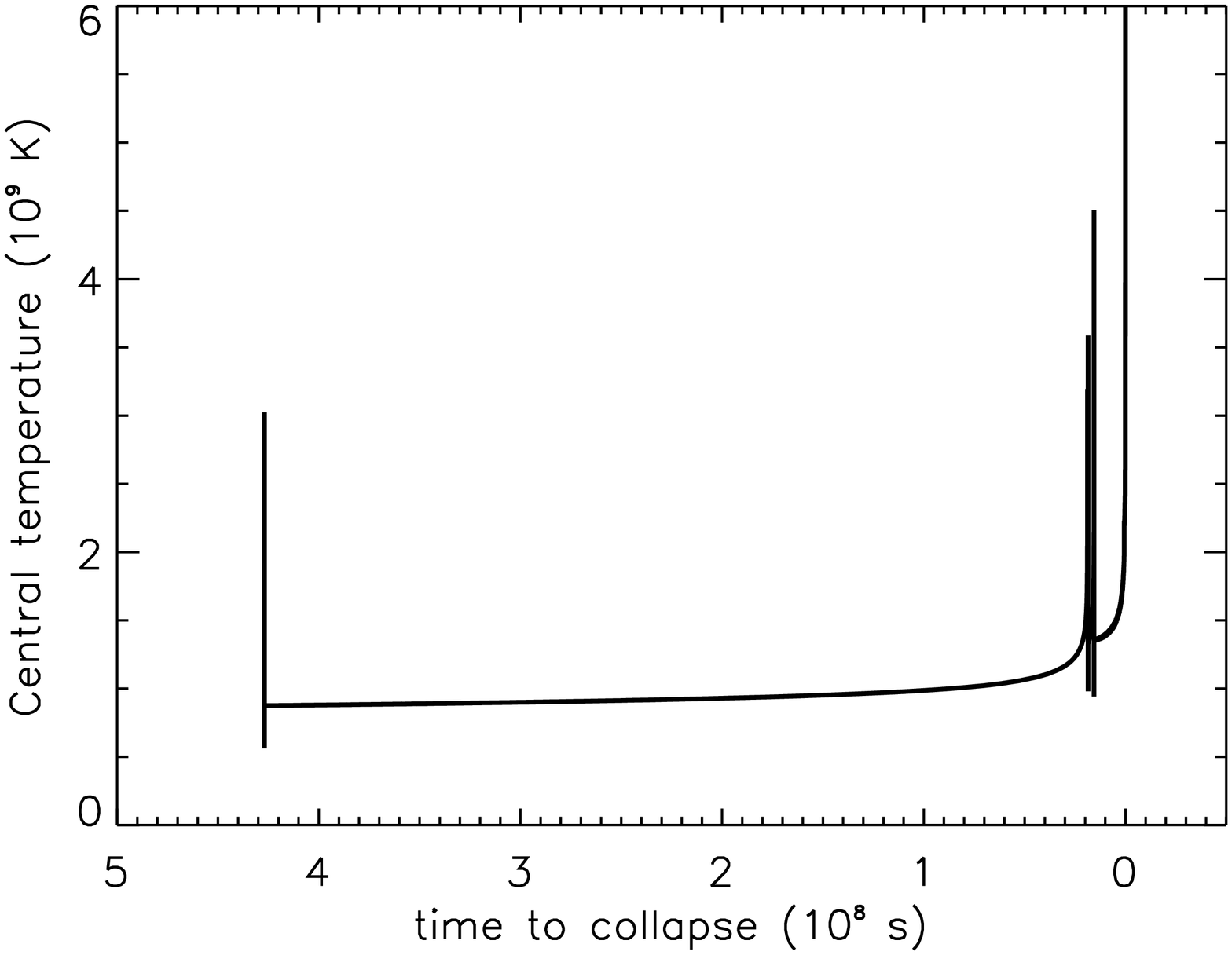} \\
\includegraphics[scale=.22]{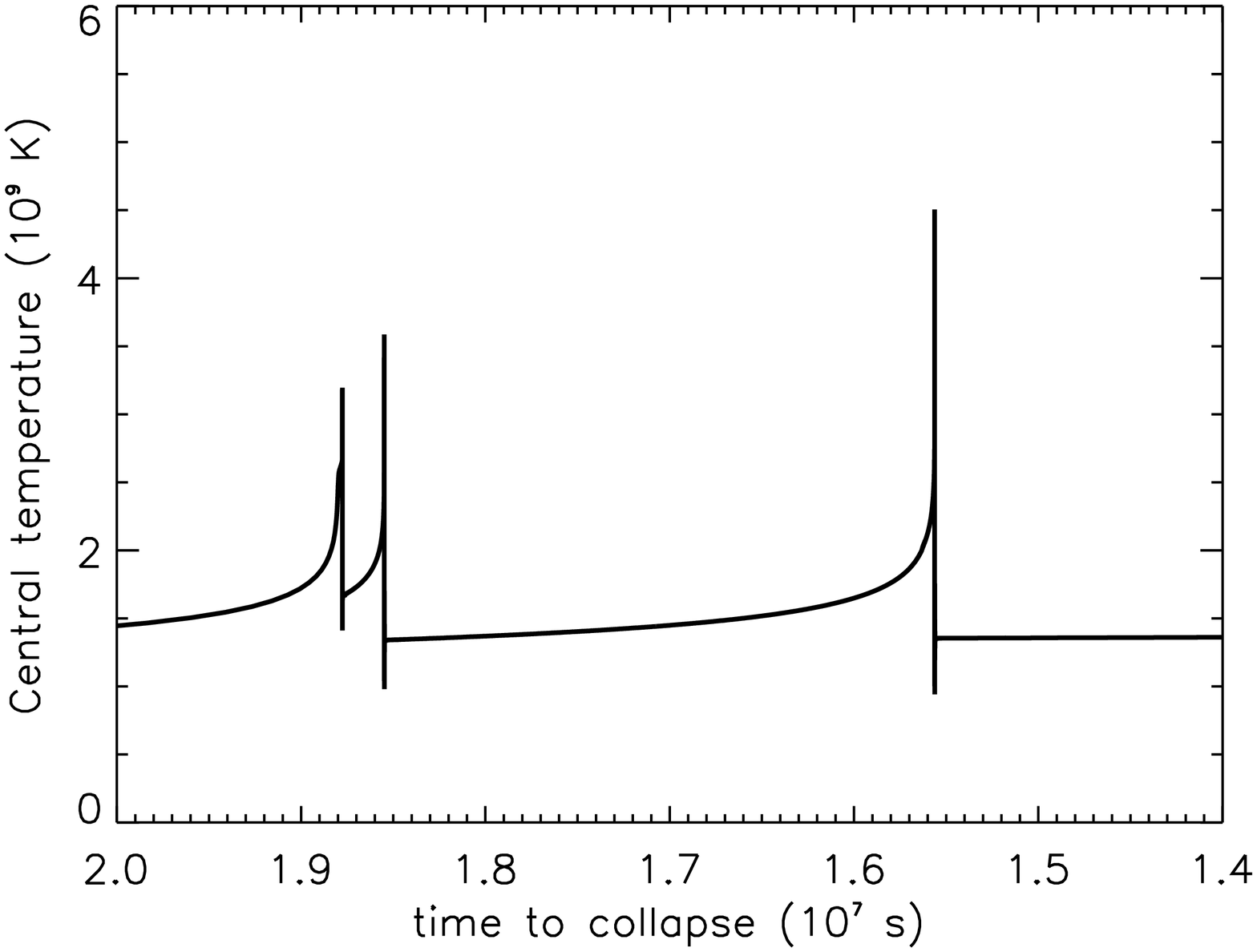} &
\includegraphics[scale=.22]{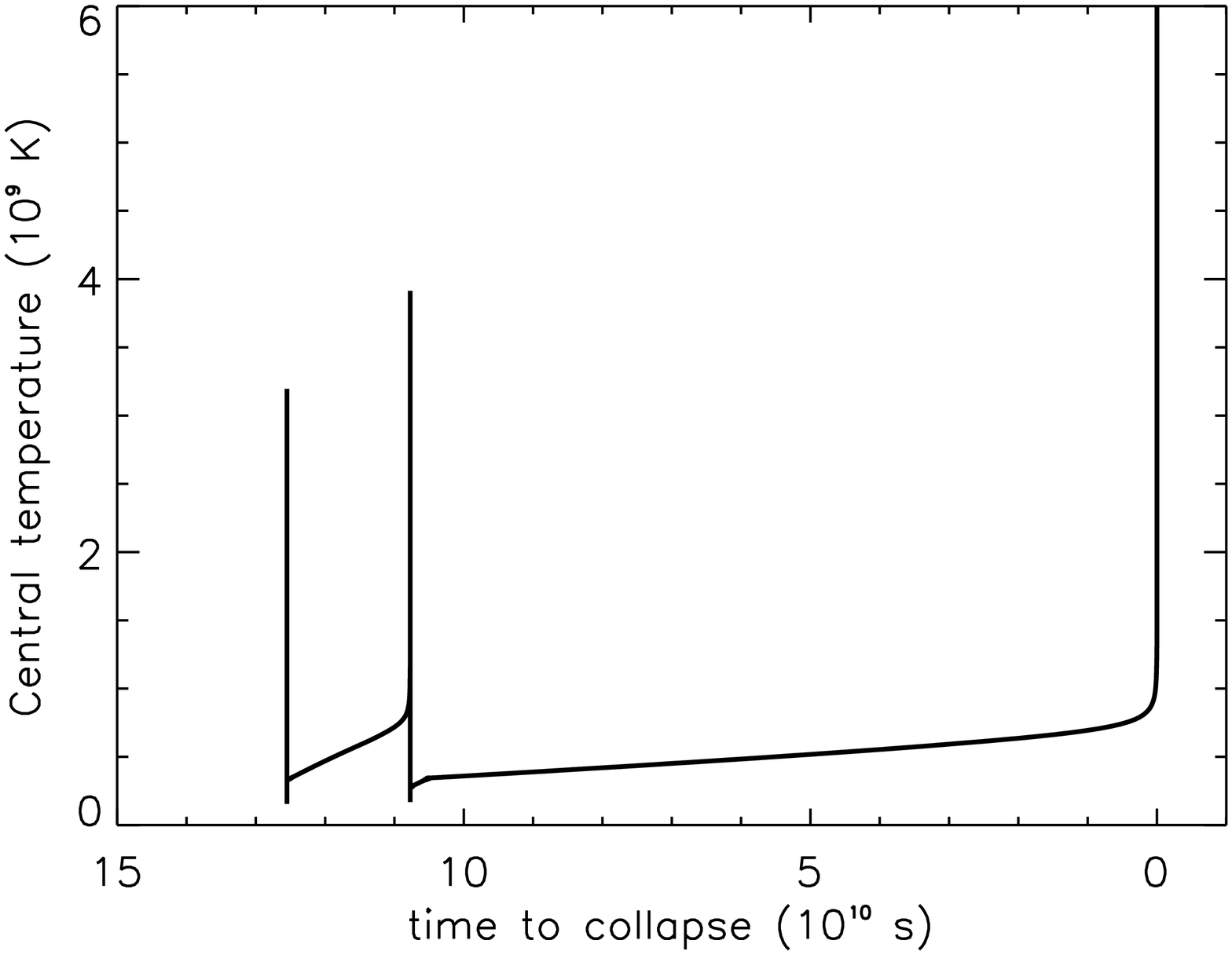} \\
\end{tabular}
\caption{Pair-driven pulsations cause rapid variations in the central
  temperature (10$^9$ K) near the time of death for helium cores of
  32, 36, 40, 44, 48, 52 (on two different time scales) and 56 \Msun
  \ (left to right; top to bottom). The log base 10 of the time scales
  (s) in each panel are respectively 4, 4, 5, 5, 6, 8, 7, and 10. The
  last rise to high temperature marks the collapse of the iron core to
  a compact object. More massive cores have fewer, less frequent, but
  more energetic pulses. All plots begin at central carbon depletion.}
\label{fig:pulses}      
\end{figure}

While the observable display is quite sensitive to whether the
presupernova star retains its hydrogen envelope or not, the number,
energies, and duration of the pulses driven by the pair instability is
determined entirely by the helium core mass. One can thus sample the
broad properties of PPSN using only a grid of bare helium cores. This
has the appealing simplicity of removing the uncertain effects of
convective dredge up and rotational mixing during hydrogen burning and
reducing the problem to a one parameter family of outcomes.
Table~\ref{tab:ppsntab} and Figure~\ref{fig:pulses} summarize some
recent results for helium cores of various masses.

Initially, the instability is quite mild and only happens very close
to the end of the star's life, after it has already completed core
oxygen burning and is burning oxygen in a shell. For larger helium
core masses, a few pulses contribute sufficient energy (about
10$^{48}$ erg), that starting at around 34 \Msun, the hydrogen
envelope is ejected, but little else. The low energy ejection of the
envelope produces very faint, long lasting Type IIp
supernovae. The continued evolution of such stars yields an iron core
of about 2.5 \Msun \ that almost certainly collapses to a black hole
with a mass nearly equal to the helium core mass. Thus the ejection of
the envelope and its nucleosynthesis are the only observables for a
distant event.

\begin{table}
\caption{Pulses from Helium Core Explosions of Different Masses (\Msun)}
\label{tab:ppsntab}       
\begin{tabular}{p{2cm}p{2cm}p{2cm}p{2cm}p{2cm}}
\hline\noalign{\smallskip}
Mass & N Pulse & Duration & Energy & Rem. Mass  \\
\noalign{\smallskip}\svhline\noalign{\smallskip}
32 &    weak     &      4.0(3)  &   1.6(45)  &  32    \\
34 &    12       &      6.5(3)  &   1.5(48)  &  33.93 \\
36 &    many     &      1.4(4)  &   9.2(48)  &  35.81 \\
38 &    many     &       8.7(4) &   1.1(50) &  37.29 \\
40 &    many     &      2.8(5)  &   2.7(50)  &  38.24 \\
42 &    18       &      3.3(5)  &   2.4(50)  &  39.72 \\
44 &    10       &      9.0(5)  &   5.8(50)  &  39.94 \\
46 &    10       &      2.2(6)  &   6.6(50)  &  41.27 \\
48 &     7       &      6.4(6)  &   9.2(50)  &  41.52 \\
50 &     4       &      7.1(7)  &   8.1(50)  &  42.80 \\
52 &     4       &      4.3(8)  &   8.1(50)  &  45.87 \\
54 &     2       &      5.4(10) &   1.6(51)  &  43.35 \\ 
56 &     2       &      1.3(11) &   1.6(51)  &  40.61 \\ 
58 &     2       &      3.0(11) &   3.7(51)  &  17.06 \\   
60 &     2       &      1.3(11) &   2.7(51)  &  36.60 \\
62 &     2       &      5.3(11) &   7.1(51)  &   5.33 \\           
64 &     1       &        -     &   4.7(51)  &    0   \\
66 &     1       &        -     &   6.8(51)  &    0   \\
\noalign{\smallskip}\hline\noalign{\smallskip}
\end{tabular}
\end{table}

Moving on up in mass, the pulses have more energy, start earlier, and
increase in number until, above 42 \Msun, their number starts to
decline again. Figure~\ref{fig:pulses} shows that in the mass range 36
\Msun to about 44 \Msun \ a major pulse is typically preceded by a
string of smaller ones that grow in amplitude until a single violent
event causes a major change in the stellar structure. Recovery from
this violent event requires a Kelvin-Helmholtz time scale ($\tau_{\rm
  KH} \sim GM^2/RL$) for the core to contract back to the unstable
temperature, around $2 \times 10^9$ K. If the pulse is a weak one, the
luminosity in the Kelvin-Helmholtz time scale is the neutrino
luminosity and is large, making the time scale short. If the pulse
decreases the central temperature below a half-billion degrees
however, radiation transport enters in and the time scale becomes
long.  On the heavier end of this mass range, the total energy of
pulses is a few times 10$^{50}$ erg, but their overall duration is
less than a week. Since this is less than the time required for the
ejected matter to become optically thin, the collisions are usually
finished before any supernova becomes visible. Depending upon the
presence of an envelope, one expects, for these cases, a rather
typical Type Ib or IIp light curve, with some structure possible in
the case of the bare helium core because of its short shock
transversal time (Section-\ref{sec:helite}). When the pulses are over,
a large iron core is again produced, and, some time later, the
remaining core of helium and heavy elements probably becomes a black
hole.

\begin{figure}
\centering
\begin{tabular}{cc}
\includegraphics[scale=.22]{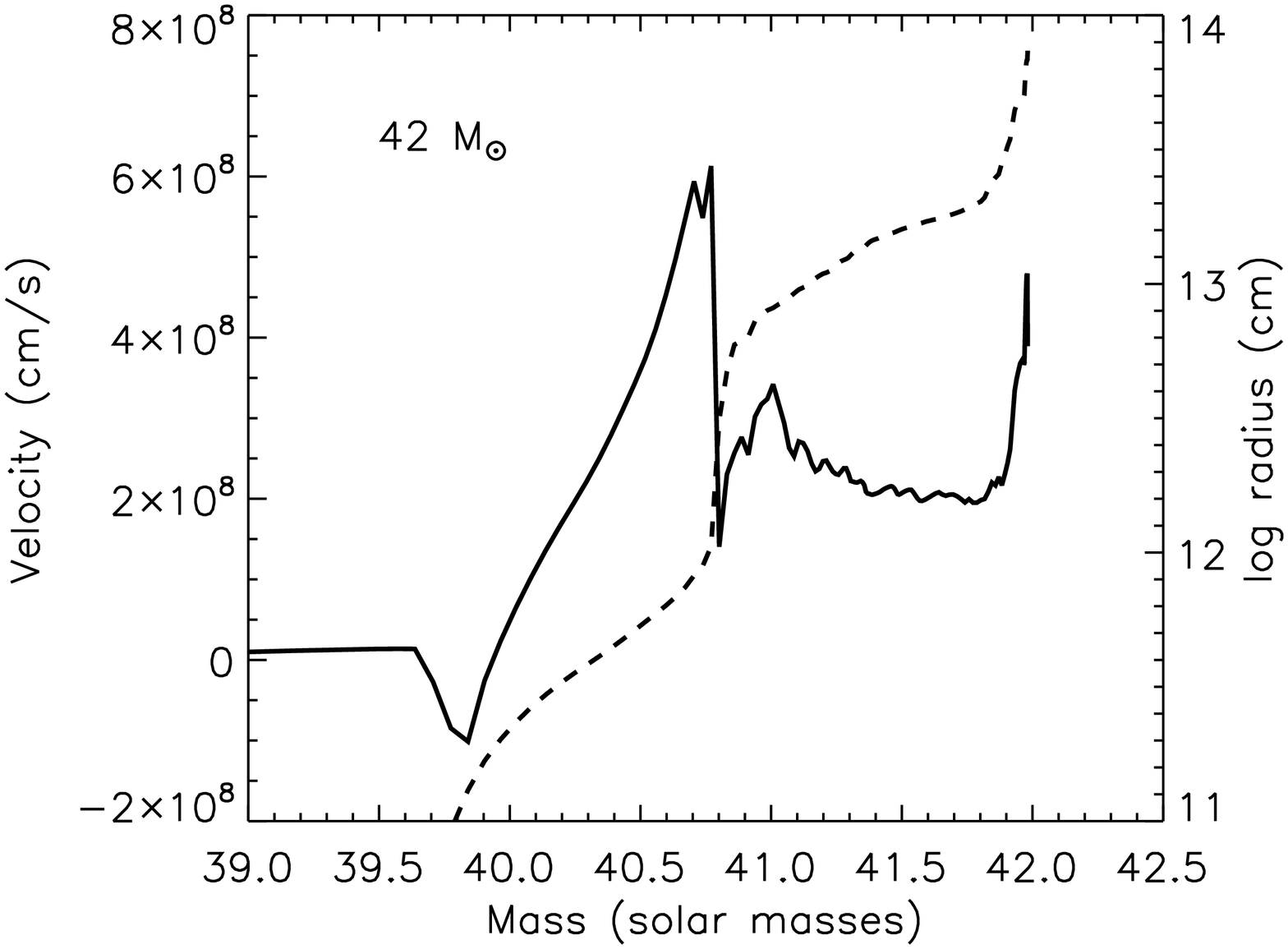} &
\includegraphics[scale=.22]{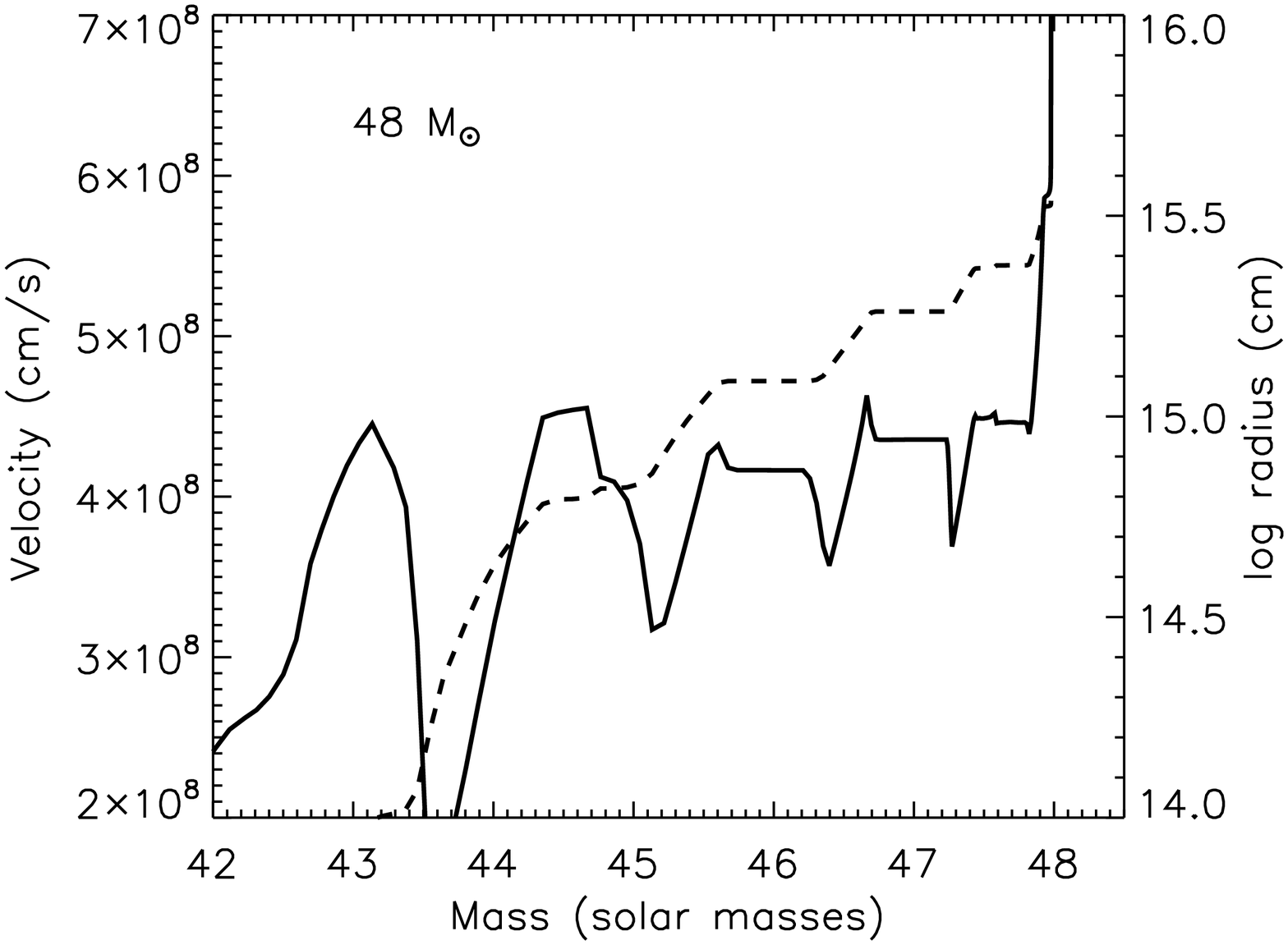} \\
\includegraphics[scale=.22]{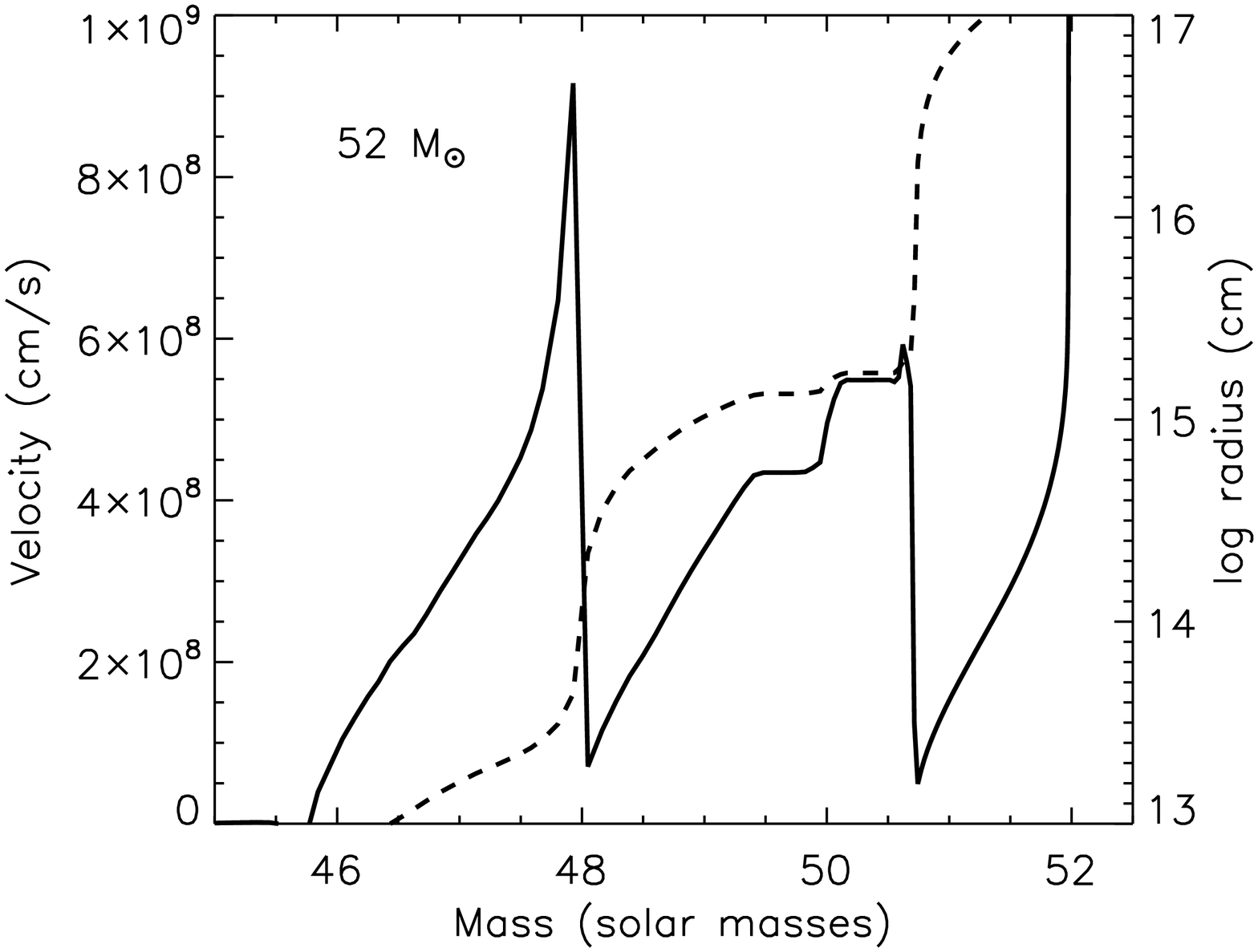} &
\includegraphics[scale=.22]{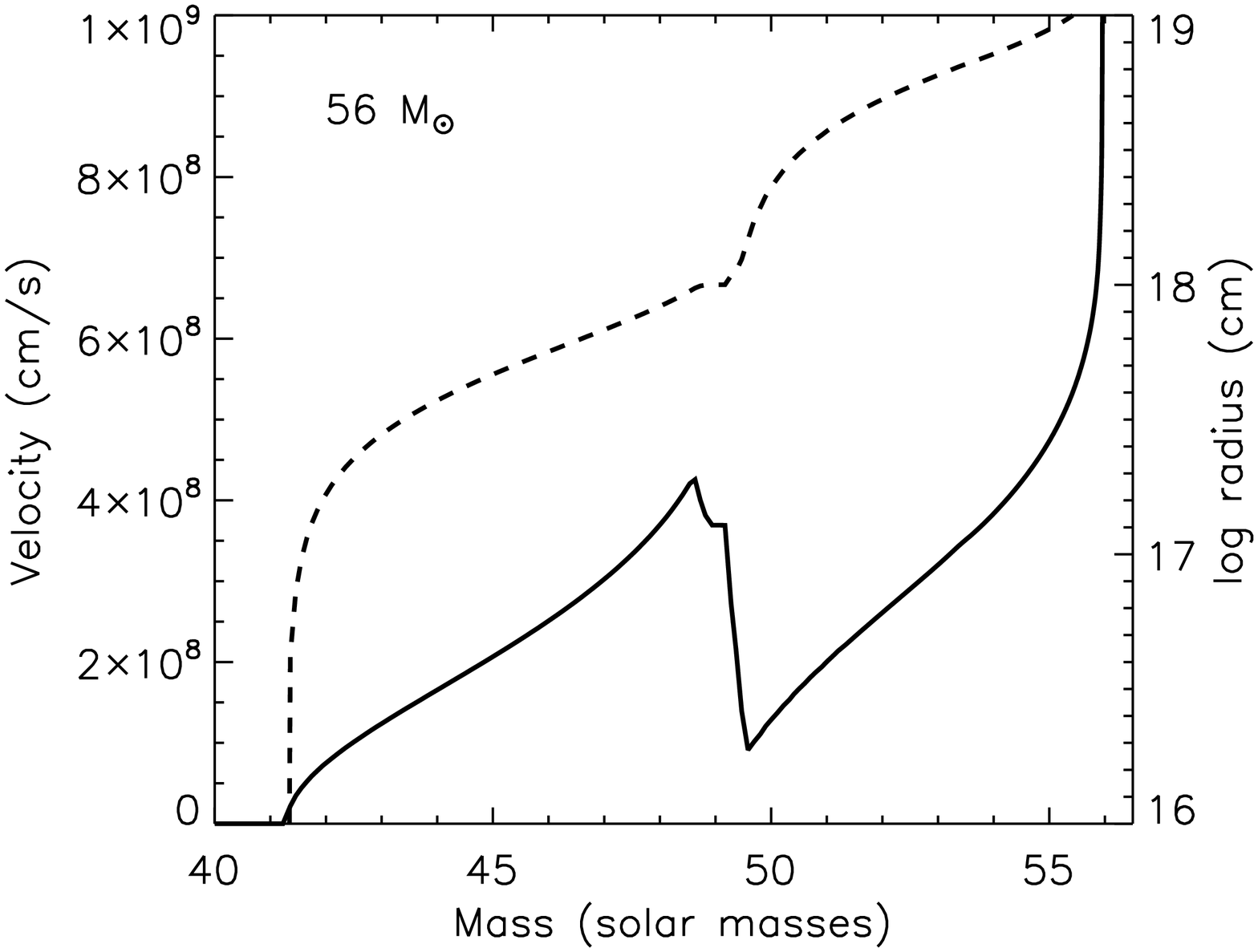} \\
\end{tabular}
\caption{Velocities (solid lines) and radii (dashed lines) of ejected
  shells for four helium cores producing mass ejection by the
  pulsational pair mechanism. The velocities are evaluated at various
  times when the collision between shells is underway. For the 42 and
  48 \Msun models. this was near iron core collapse. For 52 \Msun, it
  was at central silicon depletion, and for 56 \Msun, after a strong
  silicon flash, but before the re-ignition of silicon. Some merging of
  pulses has already occurred. Regions of flat velocity imply
  spatially thin, high density shells that may be unstable in two or
  three dimensions.}
\label{fig:ppsnvel}      
\end{figure}

For still heavier helium core masses, 44 to 52 \Msun, the total energy
of the pulses becomes that of a typical supernova, but spread over
several pulses that require from weeks to years to complete. An
important alignment of time scales occurs in this mass range. For the
masses and energies ejected, average shell speeds for the first pulse
are a few thousand km s$^{-1}$ (much less if a hydrogen envelope is in
the way). At this speed, a radius of $\sim$10$^{16}$ cm is reached in
about a year, which is comparable to the interval between
pulses. Repeated supernovae and supernovae with complex light curves
are thus possible. The photospheric radii of typical supernovae in
nature are a few times 10$^{15}$ cm, this being the distance where the
expanding debris most efficiently radiate away their trapped energy on
an explosive time scale. Since the ejecta of a given pulse will
consist of material moving both slower than and faster than the
average, and because each pulse is typically more energetic than its
predecessor, shells collide at radii 10$^{15} - 10^{16}$ cm
(Figure~\ref{fig:ppsnvel}).

These collisions convert streaming kinetic energy to optical light
with high efficiency. In principle, a substantial fraction of the
total kinetic energy of the pulses can be radiated, especially if the
shells all run into a slowly moving hydrogen shell ejected in the
first pulse. Stars in this mass range, in the most extreme cases, can
thus give repeated supernovae with up to 10$^{51}$ erg of light.

%

Still more energetic and less frequent pulsations happen at higher
mass, but now the presence of the envelope becomes critical. Without
a hydrogen envelope, the time between pulses is so long that the
collisions happen at very large radii, 10$^{17}$ - 10$^{18}$
cm. For these very large radii, the result would not be so different
from an ordinary 10$^{51}$ erg supernova running into an unusually
dense interstellar medium. Both the very large radii and long time
scales preclude any resemblance to ordinary optical supernovae, but
the events might instead present as bright radio and x-ray transients.

In the presence of an envelope, the first pulse does not eject matter
with such high speed and, given the large variation in speed from the
inner part of the moving shell to its outer extremity, substantial
energy could still be emitted by explosions in this mass range by
shells colliding inside of 10$^{16}$ cm making a bright Type II
supernova.

Pulses continue until the helium core has lost enough mass to be
stable again. This gives a range of remnant masses typically around 34
to 46 \Msun \ (Table~\ref{tab:ppsntab}).  The iron core masses and
compactness parameters for these stars are both very large, so it
seems very likely that black holes will result for the entire range of
stars making PPSN, all having typical masses around 40 \Msun.

\subsection{Light Curves for Helium Stars}
\label{sec:helite}

\begin{figure}
\centering
\begin{tabular}{cc}
\includegraphics[scale=.22]{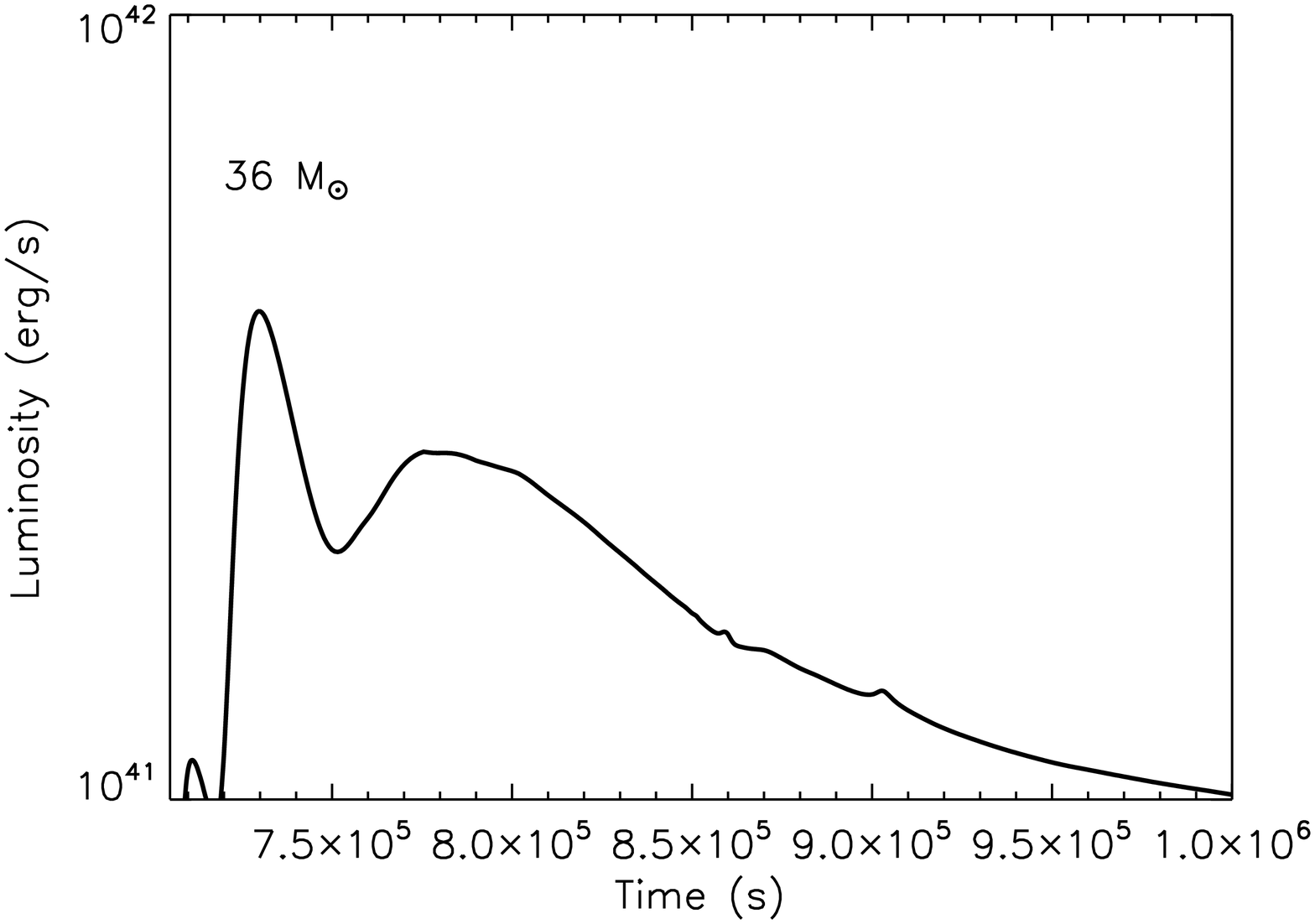} &
\includegraphics[scale=.22]{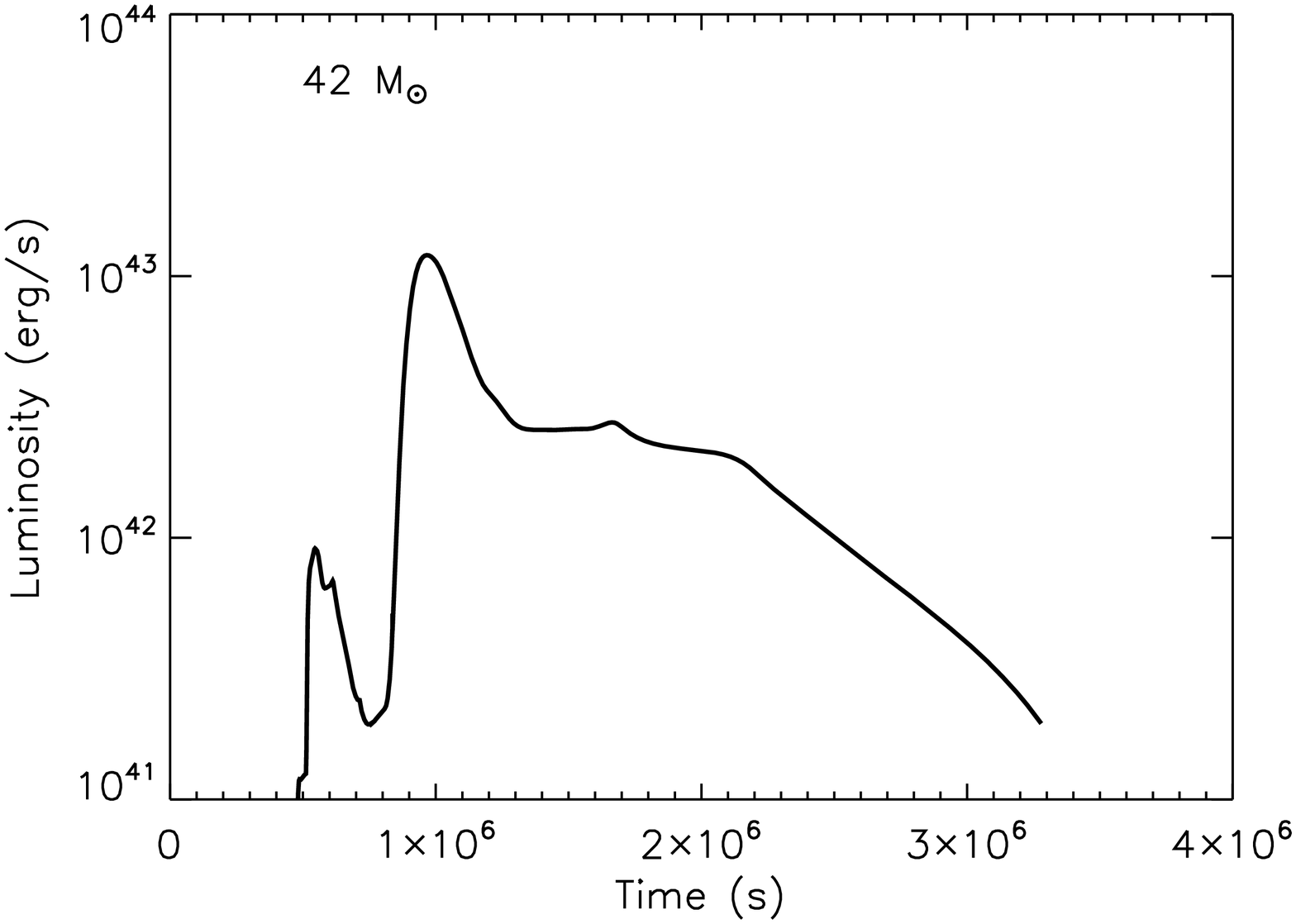} \\
\includegraphics[scale=.22]{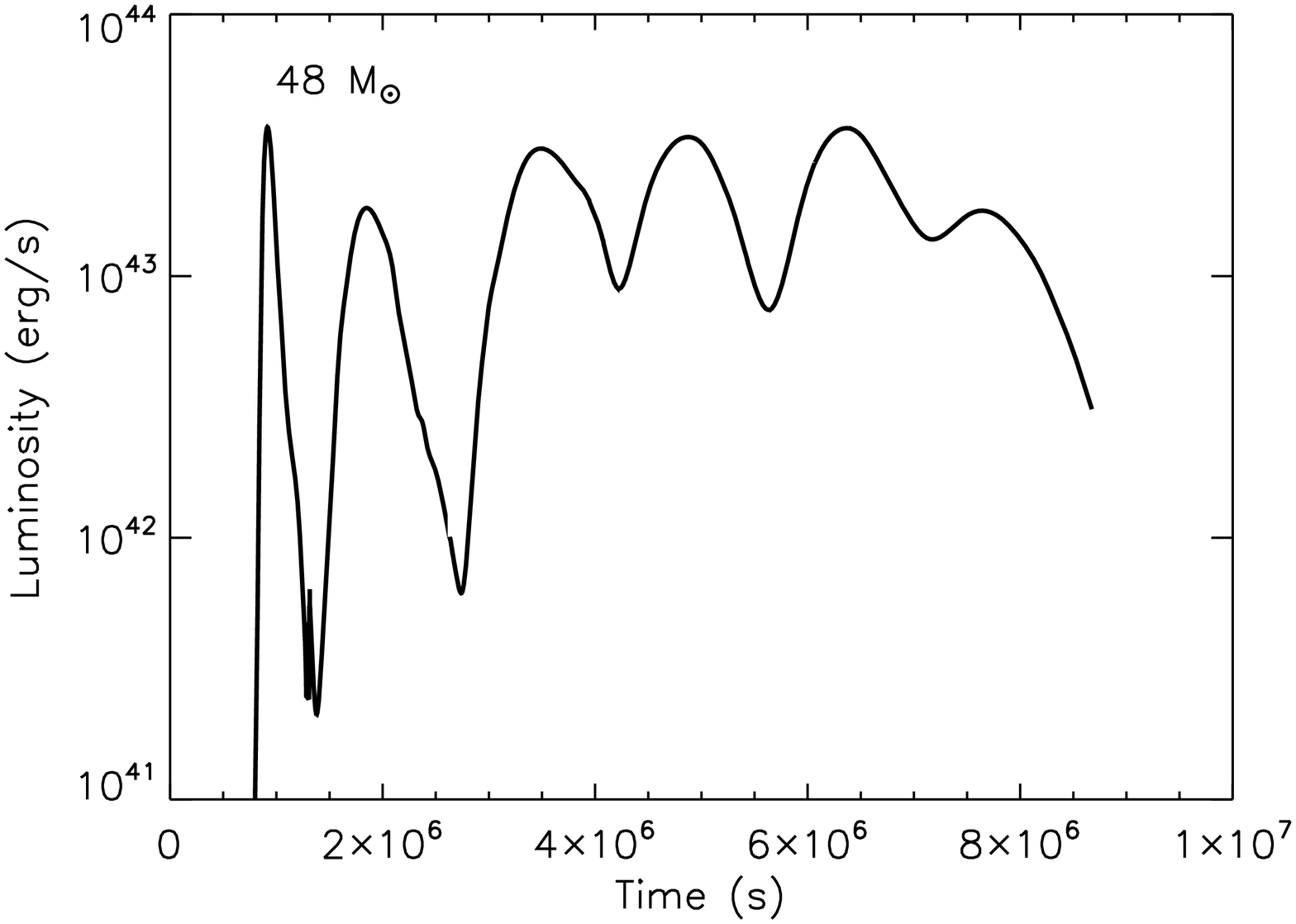} &
\includegraphics[scale=.22]{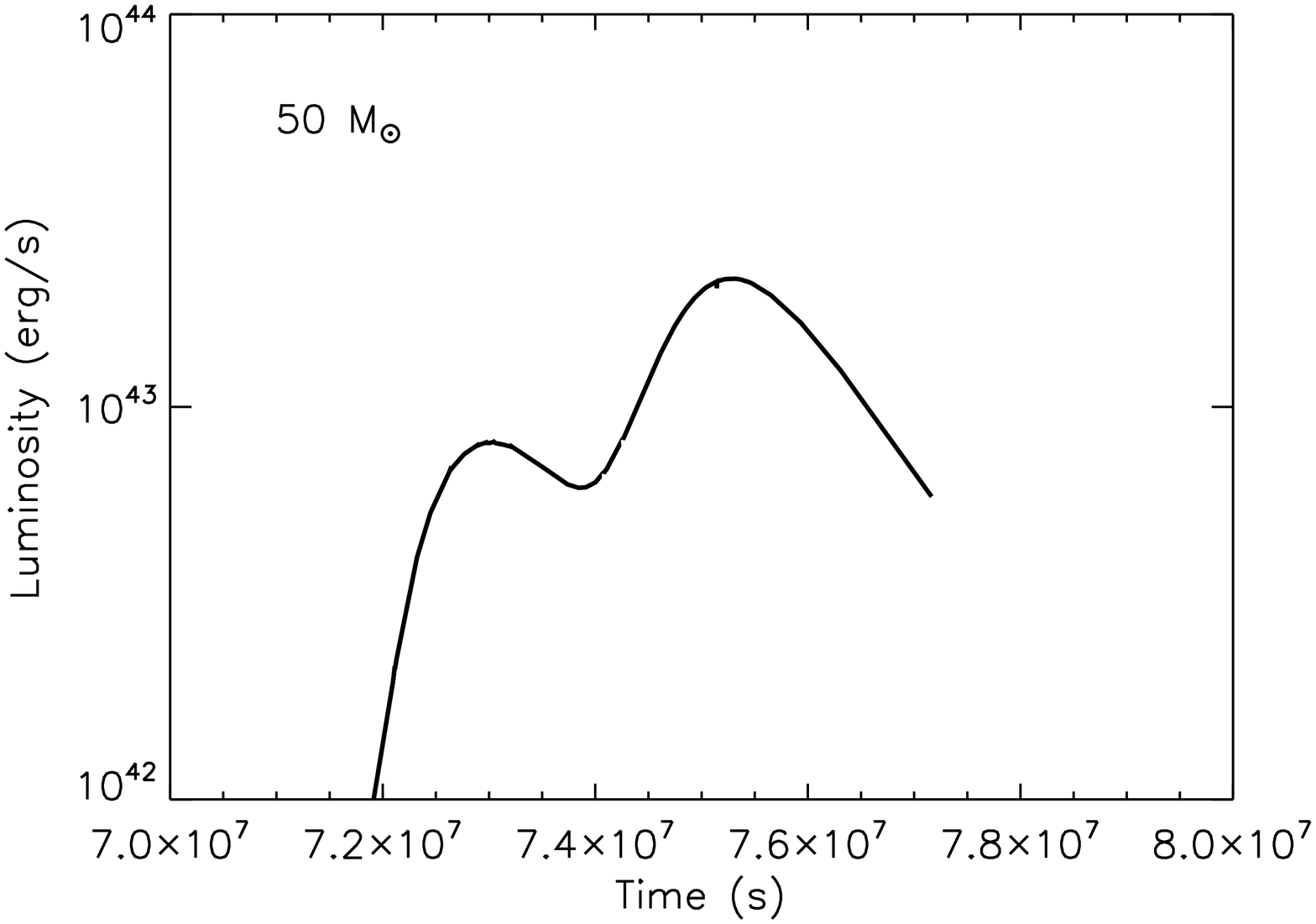} \\
\end{tabular}
\caption{Bolometric light curves from pulsational pair instability
  supernovae derived from bare helium cores of 36, 42, 48 and 50
  \Msun. A wide variety of outcomes is possible. For the 36 and 42
  \Msun \ models the photospheric radius is inside 10$^{15}$ cm and
  the transients will be blue. For the higher two masses, the
  photosphere is near 10$^{15}$ cm and the transients might have
  colors more like an ordinary supernova.}
\label{fig:ppsnlite}      
\end{figure}

Light curves for a sample of helium core explosions are shown in
Figure~\ref{fig:ppsnlite} and illustrate the characteristics discussed
in the previous section. For the lighter helium cores, the pulses only
eject a small amount of matter with low energy. Shell collisions are
over before light escapes from the collision region. The light curve
for the 26 \Msun \ helium core is typical for this mass range - a
subluminous ``supernova'' of less than 10$^{42}$ erg s$^{-1}$ lasting
only a few days. These might be looked for in the case of stars that
have lost their envelopes prior to exploding. In a star with an
envelope, as we shall see later, the situation would be very
different. Even the small (10$^{49}$ erg) kinetic energy would unbind
the envelope producing a long, faint Type IIp supernova.

For the 42\Msun \ helium core, a brighter, longer lasting transient is
produced, but still only a single event, albeit a structured one. The
total duration of pulses is about 2 days, followed by a 2 day wait
until the core collapse. The last pulse is a particularly violent one.
The light curve (Figure~\ref{fig:ppsnlite}) shows a faint
outburst occurring as many smaller pulses merge and the first big
of mass is ejected, followed by a longer more luminous peak as that
main pulse runs into the prior ejecta. Both of these transients are
quite blue since the collisions are occurring at small radius, a few
times 10$^{14}$cm.

By 48 \Msun, the shell collisions are becoming sufficiently energetic
and infrequent that the light curve fractures into multiple
events. The collisions are now happening at around 10$^{15}$ cm and
should be quite bright optically. At 52 \Msun, one sees repeated
individual supernovae.  Figure~\ref{fig:ppsnlite} merely shows the
brightest one from this object.  Activity at the 10$^{41}$ erg level
started two years before.

It should be noted, though, that all these 1D light-curve calculations
are quite approximate and need to be repeated in a multi-dimensional
code with the appropriate physics, especially for cases where the
shells collide in an optically thin regime. KEPLER, a one dimensional
implicit hydrodynamics code with flux-limited radiative diffusion does
an admirable job in a difficult situation. In 1D however, the
snowplowing of a fast-moving shell into a slower one generates a large
spike in density, with variations of many orders of magnitude in
density between one zone and an adjacent one. For a time this thin
shell corresponds to the photosphere. The ``linearized'' equations of
hydrodynamics do not behave well in such clearly non-linear
circumstances and the outcome of a multi-dimensional calculation may
be qualitatively different. This is an area of active research.

\subsection{Type II Pulsational Pair Instability Supernovae}
\label{sec:ppsnii}

The retention of even a small part of the original hydrogen envelope
significantly alters the dynamics and appearance of PPSN. For example,
what wold have been a brief, faint transient for a 36 \Msun
\ helium core (Figure~\ref{fig:ppsnlite}), provides more than enough
energy to eject the entire envelope of a red supergiant. A great
diversity of outcomes is possible depending upon the mass of the
envelope and helium core and the radius of the envelope

\begin{figure}
\centering
\begin{tabular}{cc}
\includegraphics[scale=.6]{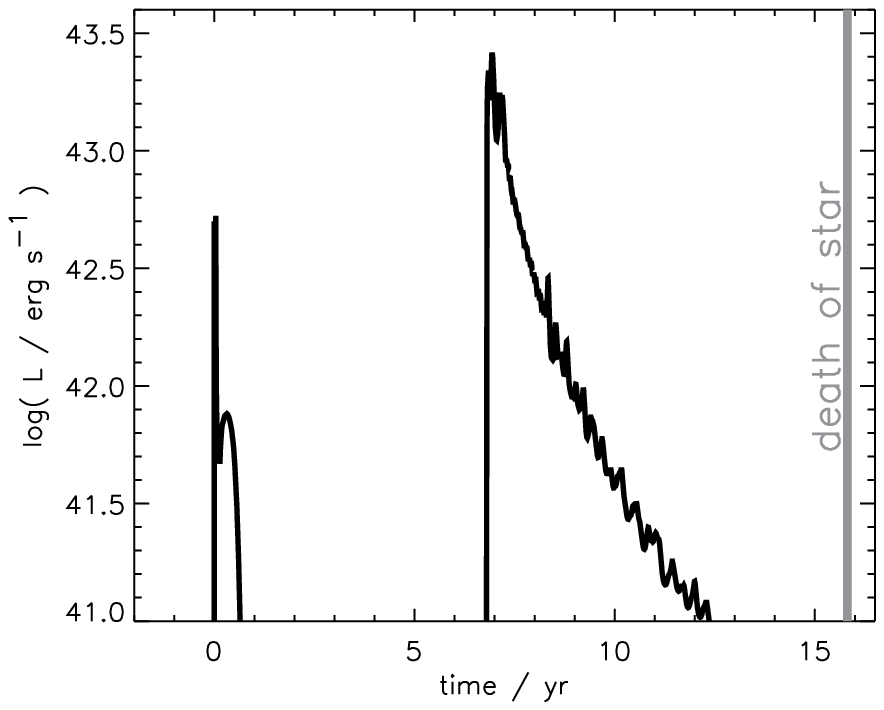} &
\includegraphics[scale=.6]{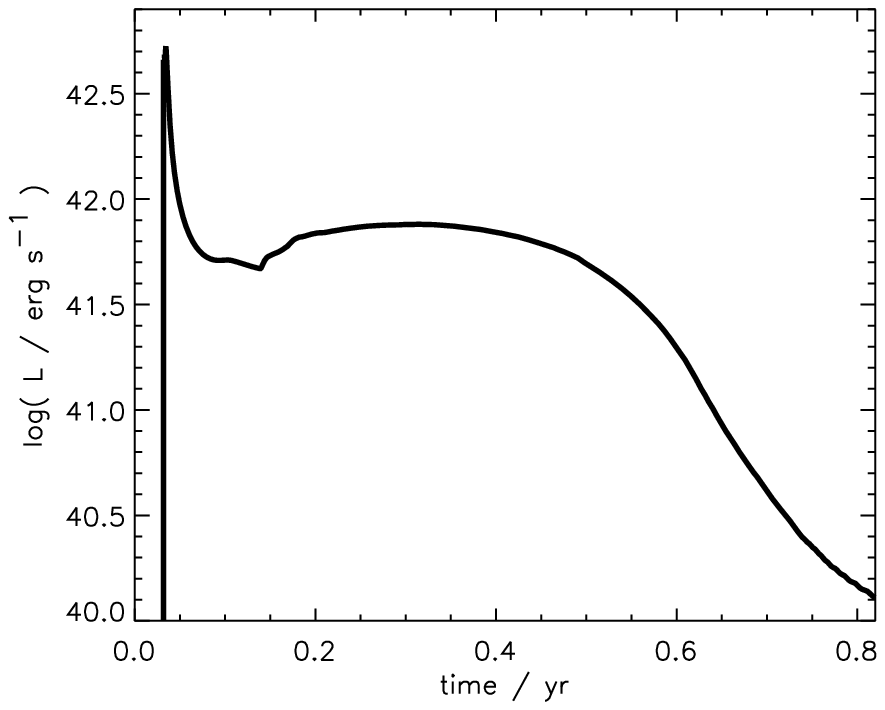} \\
\end{tabular}
\caption{Light curves of the two supernovae produced by the 110 \Msun
  \ PPSN \citep{Woo07}. The first pulse ejects the envelope and
  produces the faint supernova shown in greater detail on the
  right. 6.8 years later the collision of pulses 2 and 3 with that
  envelope produces another brighter outburst (see
  Figure~\ref{fig:ppsn2d})}
\label{fig:ppsnlc}      
\end{figure}

\begin{figure}
\centering
\begin{tabular}{cc}
\includegraphics[scale=.33]{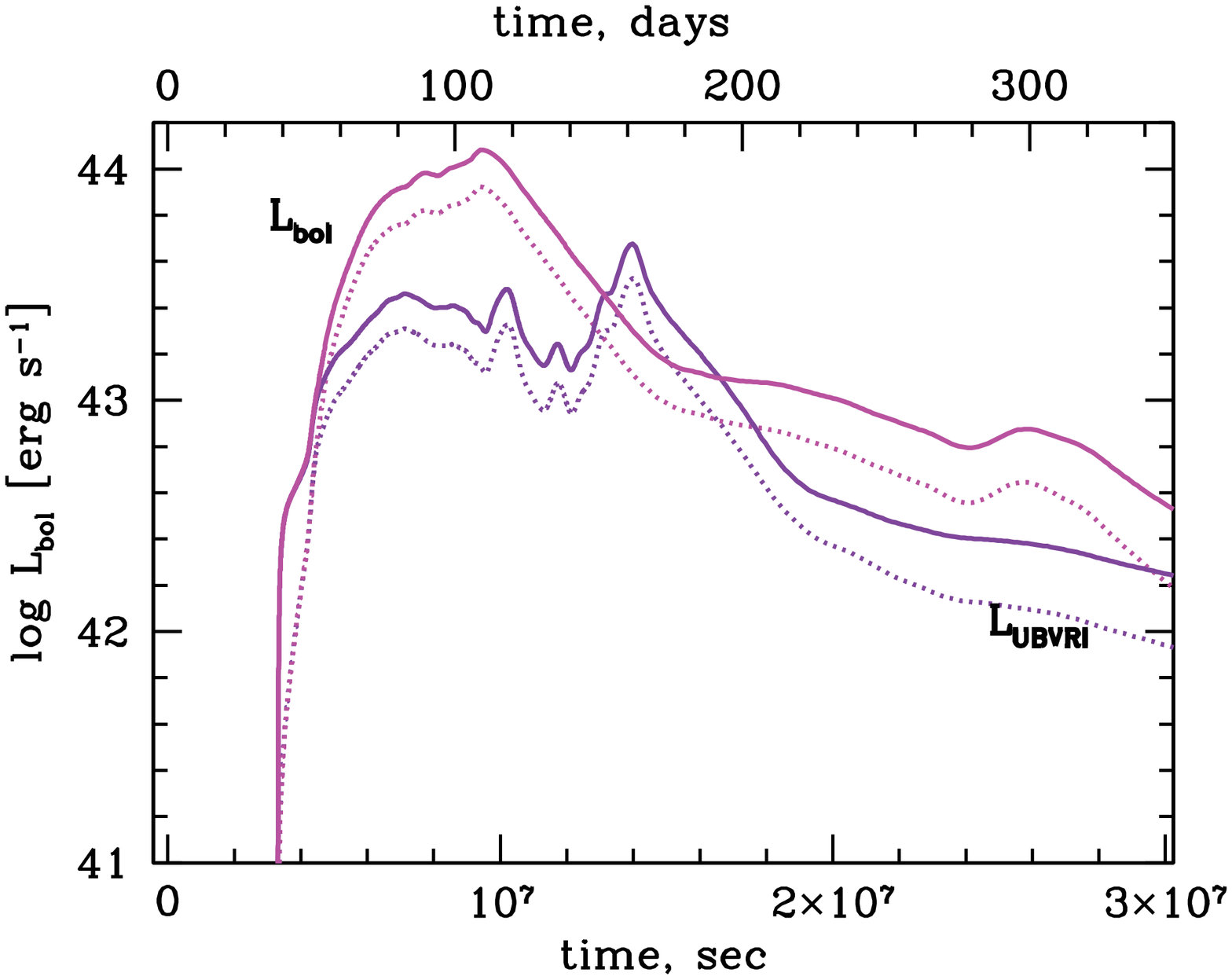} &
\includegraphics[scale=.18]{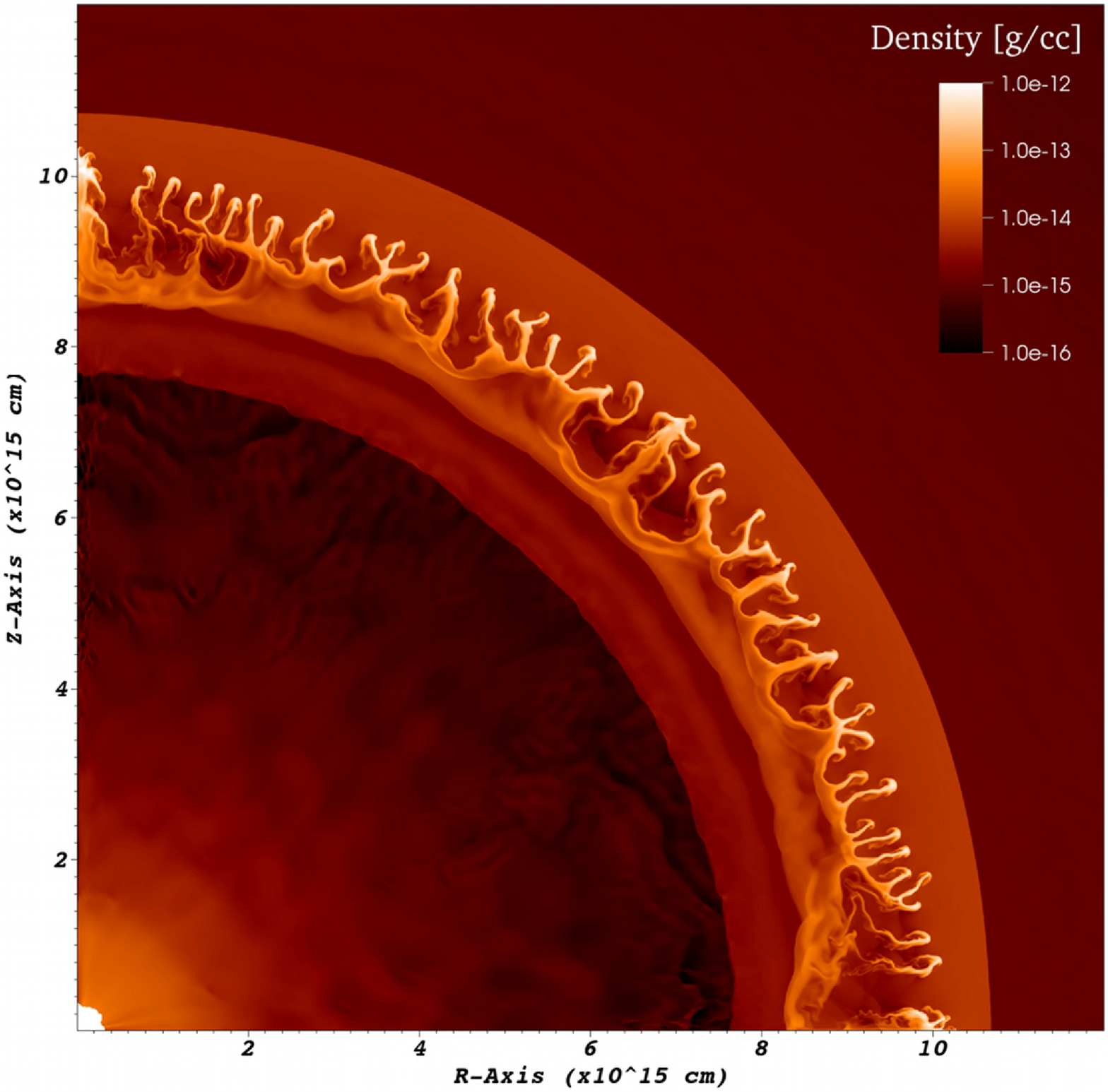} \\
\end{tabular}
\caption{{\sl Left:} Light curve for the second very luminous outburst
  of the 110 \Msun \ model (see Figure~\ref{fig:ppsnlc}) of the two
  supernovae produced by the 110 \Msun \ PPSN \citep{Woo07}.  The
  brighter set of curves results hen the collision speed is
  artificially increased by a factor of 2 and resembles SN
  2006gy. {\sl Right:} 2D calculation of the explosion of a 110 \Msun
  \ star as a PPSN. The dense shell produced in 1D by the collision of
  the ejecta from two pulse is Rayleigh-Taylor unstable. The resulting
  density contrast is much smaller.}
\label{fig:ppsn2d}      
\end{figure}

Most striking are the ``ultra-luminous supernovae'' of Type IIn that
happen when very energetic pulses from the edge of the helium core
strike a slowly moving, previously ejected hydrogen envelope. A
similar (Type I) phenomenon could happen for bare helium cores, but
probably with a shorter-lived, less luminous light curve owing
to the smaller masses involved. An example is shown in
Figure~\ref{fig:ppsn2d} based upon the evolution of a 110 \Msun \ star
\citep{Woo07}. By the end of its life this star had shrunk to 74.6
\Msun \ (using a wholly artificial mass loss rate), of which 49.9
\Msun \ was the helium core. This core experienced three violent
pulsations. The first ejected almost all of the hydrogen envelope,
leaving 50.7 \Msun \ behind. This envelope ejection produced a rather
typical Type IIp supernova although with a slower than typical speed
and luminosity (Figure~\ref{fig:ppsnlc}). By 6.8 years later, the
stellar remnant had contracted to the point that it experienced the
pair instability again. Two more pulses, occurring in rapid succession,
ejected an additional 5.1 \Msun \ with a total kinetic energy of $6
\times 10^{50}$ erg. Pulses 2 and 3 quickly merged and then run into
the ejected envelope (Figures~\ref{fig:ppsnlc} and \ref{fig:ppsn2d}).

These light curves were calculated using 1D codes in which the
collision of the shells again produced a very large density spike. When the
calculation was run again in 2D, but without radiation transport
(Figure~\ref{fig:ppsn2d}), a Rayleigh-Taylor instability developed
that led to mixing and a greatly reduced density contrast. The
combined calculation of multi-D hydro coupled to radiation transport
has yet to be carried out, so the light curves shown here are to be
used with caution, but a multi-dimensional study would probably give a
smoother light curve.

\subsection{Nucleosynthesis}

The nucleosynthesis from PPSN is novel in that it is heavily weighted
towards the light species that are ejected in the shells. For present
purposes, given the large iron cores, we assume that all matter not
ejected by the pulsations becomes a black hole. This assumption could
be violated if rapid rotation energized some sort of jet-like outflows
(e.g., a gamma-ray burst), but otherwise it seems reasonable.

Table~\ref{tab:ppsnnuc} gives the approximate bulk nucleosynthesis, in
solar masses, calculated for our standard set of helium cores models.
For the lightest cores, the pulses lack sufficient energy to eject
more than a small amount of surface material, which by assumption here
is pure helium. It should be noted, however, that even these weak
explosions would eject at least part of the hydrogen envelope of any
red supergiant (typical binding energy less than 10$^{48}$ erg). Since
these envelopes often produce primary nitrogen by mixing between the
helium core and hydrogen burning shell, an uncertain but possibly
large yield of carbon, nitrogen, and oxygen (and of course hydrogen
and helium) would accompany these explosions in a star that had not
lost its envelope.

Moving up in mass, the violence of the pulses increases rapidly and
more material is ejected, eventually reaching the deeper shells rich in
heavier elements. In Table~\ref{tab:ppsnnuc}, total yields of less
than 0.01 \Msun \ have not been included with the single exception of
the 66 \Msun \ model which made 0.037 \Msun of $^{56}$Ni. The 64 and
66 \Msun \ models are actually full up pair instability supernovae and
leave no remnants, so perhaps including their yields here with the PPSN
is a bit misleading.

If one folds these yields with an IMF to get an overall picture of the
nucleosynthesis from a generation of PPSN, it is clear that the
production (and the typical spectra of PPSN) will be dominated by H,
(He), C, N, O, (Ne) and Mg and little else. In particular, PPSN make
no iron-group elements. Given the dearth of strong He and Ne lines,
one might expect that the generation of stars following a putative
``first generation'' of PPSN would show enhancements of C, N, O, and
Mg and be ``ultra-iron poor''. Of course {\sl some} heavier elements could
be made by stars sufficiently light (main sequence mass less than 20
\Msun?) to explode by the neutrino-transport process, or sufficiently
heavy to make iron in a pair-instability supernova (helium core mass
over 65 \Msun).

\begin{table}
\caption{Nucleosynthesis in Ejected Shells (\Msun) from Helium Core Pulsational Explosions}
\label{tab:ppsnnuc}       
\begin{tabular}{p{1cm}p{1cm}p{1cm}p{1cm}p{1cm}p{1cm}p{1cm}p{1cm}p{1cm}p{1cm}p{1cm}}
\hline\noalign{\smallskip}
Mass & Total & He  & C  &  O  &  Ne  &  Mg  &  Si  & S  & Ar & Ca \\
\noalign{\smallskip}\svhline\noalign{\smallskip}
34 &  0.071& 0.071 &  -   &  -   &  -    &  -   & -    &    -  & -  & -      \\
36 &  0.19 &  0.19 &  -   &  -   &  -    &  -   & -    &    -  & -  & -      \\
38 &  0.71 &  0.32 & 0095 & 0.17 & 0.096 & 0.032& -    &    -  & -  & -      \\
40 &  1.76 &  0.50 & 0.29 & 0.53 & 0.32  & 0.11 & -    &    -  & -  & -      \\
42 &  2.28 &  0.60 & 0.43 & 0.70 & 0.41  & 0.14 & -    &    -  & -  & -      \\
44 &  4.06 &  0.85 & 0.79 & 1.36 & 0.80  & 0.26 & -    &    -  & -  & -      \\
46 &  4.73 &  1.02 & 0.94 & 1.61 & 0.90  & 0.27 & -    &    -  & -  & -      \\
48 &  6.48 &  1.34 & 1.40 & 2.30 & 1.15  & 0.30 & -    &    -  & -  & -      \\
50 &  7.20 &  1.58 & 1.60 & 2.61 & 1.16  & 0.26 & -    &    -  & -  & -      \\
52 &  6.13 &  1.55 & 1.33 & 2.29 & 0.81  & 0.16 & 0.001&    -  & -  & -      \\
54 & 10.64 &  1.65 & 1.83 & 5.32 & 1.35  & 0.41 & 0.074&    -  & -  & -      \\
56 & 15.38 &  1.74 & 2.06 & 9.41 & 1.52  & 0.50 & 0.15 &    -  & -  & -      \\
58 & 40.93 &  1.85 & 2.87 & 30.5 & 2.64  & 1.42 & 1.49 &   0.17& 0.020  & 0.015      \\
60 & 23.39 &  1.89 & 3.10 & 15.0 & 2.49  & 0.60 & 0.28 & 0.058 & 0.008 & 0.005  \\
62 & 56.67 &  1.95 & 2.87 & 37.5 & 2.60  & 1.43 & 6.39 &   2.99& 0.51 & 0.44 \\
64 &   64  &  1.92 & 3.62 & 44.1 & 3.60  & 2.12 & 5.35 &   2.41& 0.43 & 0.38 \\
66 &   66  &  1.79 & 3.60 & 42.8 & 3.99  & 2.07 & 7.11 &   3.49& 0.60 & 0.53 \\
\noalign{\smallskip}\hline\noalign{\smallskip}
\end{tabular}
\end{table}

\section{150 to 260 \Msun; Pair Instability Supernovae}
\label{sec:pairsn}

The physics of pair instability supernovae (PISN) is sufficiently well
understood that they can be accurately modeled in 1D on a desktop
computer. A major question though is their frequency in the
universe. PISN come from a range of masses somewhat heavier than we
expect for presupernova stars today. This is not to say that stars of
over 150 \Msun \ are not being born. See e.g., the review by Crowther
reported in \citet{Vin13} which gives 320 \Msun \ as the current
observational limit. The issue is whether such large masses can be
retained in a star whose luminosity hovers near the Eddington limit
\citep{Vin11}. 
Still observers claim to have discovered at least one PISN event
\citep{Gal09}. Because the critical quantity governing whether a star
becomes PISN is the helium core mass of the presupernova star (greater
than 65 \Msun), they are favored by diminished mass loss, i.e., at low
metallicity, and may have been more abundant in the early universe.

\begin{figure}
\includegraphics[scale=.90]{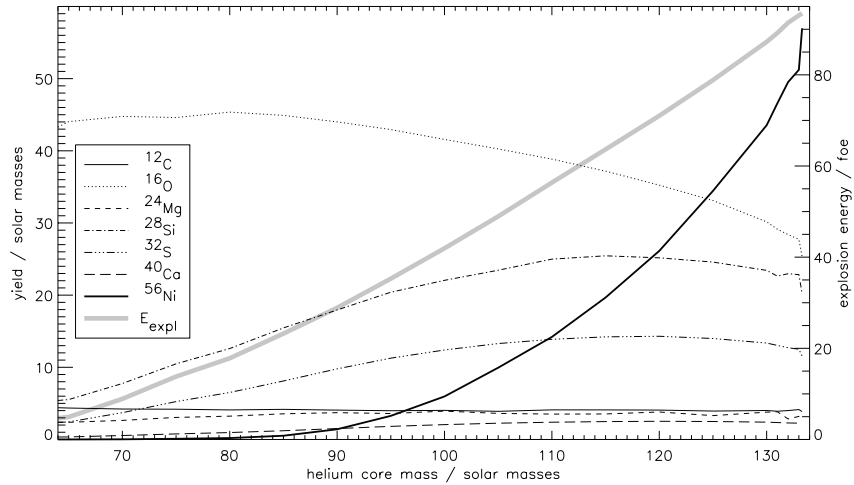} 
\caption{Nucleosynthesis in pair-instability supernovae as a function
  of helium core mass. Also given is the explosion energy in units of
  10$^{51}$ erg (broad grey line) which rises steadily with mass. The
  dark solid line is $^{56}$Ni synthesis which is not particularly
  large below 90 \Msun\ \citep{Heg02}.}
\label{fig:pisn}      
\end{figure}

A common misconception is that all PISN make a lot of $^{56}$Ni and
therefore are always very bright. As Figure~\ref{fig:pisn} shows, large
$^{56}$Ni production and very high kinetic energies are limited to a
fairly narrow range of exceptionally heavy and rare PISN. Most events
will either present as a particularly energetic Type IIp supernova or
a {\sl subluminous} SN I. For an appreciable range of masses, less
$^{56}$Ni is produced than in, e.g., a SN Ia (about 0.7 \Msun).

The nucleosynthesis of very low metallicity PISN is quite distinctive
because they lack the excess neutrons needed to make odd-Z elements
during the explosion. This is because the initial metallicity of the
star, mostly CNO, is turned into $^{14}$N during hydrogen
burning. During helium burning, $^{14}$N captures an alpha particle
experiencing a weak decay to make $^{18}$O which has two extra
neutrons. Subsequent burning stages rearrange these neutrons using
them to make isotopes and elements that require an excess of neutrons
over protons, like almost all odd Z elements do. During the collapse
phase, the time is too short for additional weak interactions so the
ejected matter ends up deficient in things like Na, Al, P, Cl, K, Sc,
V, and Mn. Very metal poor stars show no such anomalies and this
suggests that the contribution of PISN to very early nucleosynthesis
was small.

\section{Above 260 \Msun}

Stars heavier than 260 \Msun, or more specifically non-rotating helium
cores greater than 133 \Msun, are expected to produce black holes, at
least up to about 10$^5$ \Msun. Starting around 10$^5$ \Msun,
hydrogenic stars encounter a post-Newtonian instability on the main
sequence and collapse \citep{Fow64}. If these stars have near solar
metallicity (above Z = 0.005) then titanic explosions of 10$^{56}$ -
10$^{57}$ erg, powered by explosive hydrogen burning, can result for
masses in the range 10$^5$ - 10$^6$ \Msun \ \citep{Ful86}.  Lacking a large
initial concentration of CNO, stars in this mass range, collapse to
black holes.

For lighter stars, $\sim$10$^3$ - 10$^5$ \Msun, hydrogen burns stably,
but helium burning encounters the pair instability, and on
the upper end, the post-Newtonian instability. Again black hole
formation seems the most likely outcome, though this mass range has
not been fully explored.

\section{The Effects of Rotation}
\label{sec:rotate}

Rotation alters stellar evolution in two major ways. During
presupernova evolution it leads to additional mixing processes that
can stir up either regions of the star or the whole star. Generally
the helium cores of rotating stars are larger and, since the
nucleosynthesis and explosion physics of massive stars depends
sensitively upon the helium core mass, the outcome of a smaller mass
main sequence star with rotation can resemble that of a larger one
without rotation. The mixing can also increase the lifetime of the
star and its luminosity and bring abundances to the surface that might
have otherwise remained hidden. In extreme cases, rotation can even
lead to the complete mixing of the star on the main sequence, thus
avoiding the formation of a supergiant and producing a very rapidly
rotating presupernova star that might serve as a gamma-ray burst
progenitor (Section \ref{sec:grb}).

The other way rotation changes the evolution is by affecting how the
star explodes and the properties of the compact remnant it leaves
behind. Calculations that use reasonable amounts of rotation and
approximate the effects of magnetic torques in transporting angular
momentum show that rotation may play an increasingly dominant role in
the explosion as the mass of the star increases \citep{Heg05}. This is
in marked contrast to the neutrino transport model which shows the
opposite behavior (Section~\ref{sec:compact}); heavier stars are {\sl
  more} difficult to explode with neutrinos.

\begin{table}
\caption{Pulsar Rotation Rate Predicted by Models \citep{Heg05}}
\label{tab:rotate}       
\begin{tabular}{p{1cm}p{1cm}p{1cm}p{1cm}p{1cm}p{1cm}}
\hline\noalign{\smallskip}
Mass & Baryon & Gravitational  & J  &  BE  & Pulsar P  \\
(\Msun) & (\Msun) & (\Msun) & (10$^{47}$ erg s) &(10$^{53}$ erg) & (ms)  \\
\noalign{\smallskip}\svhline\noalign{\smallskip}
12 &  1.38 &  1.26 & 5.2  & 2.3 & 15   \\
15 &  1.47 &  1.33 & 7.5  & 2.5 & 11   \\
20 &  1.71 &  1.52 & 14   & 3.4 & 7.0  \\
25 &  1.88 &  1.66 & 17   & 4.1 & 6.3  \\
35 &  2.30 &  1.97 & 41   & 6.0 & 3.0  \\
\noalign{\smallskip}\hline\noalign{\smallskip}
\end{tabular}
\end{table}

Table \ref{tab:rotate} shows the expected rotation rates of pulsars
derived from the collapse of rotating stars of various main sequence
masses. The rotational energy of these neutron stars is given
approximately by 10$^{51} (5 {\rm ms}/P)^{-2}$ erg, where it is
assumed that the neutron star moment of inertia is 80 km$^2$ \Msun
\citep{Lat07}. This implies that supernova over about 20 \Msun \ or so
have enough rotational energy to potentially power a standard
supernova. Rapidly rotating stellar cores are also expected to give
birth to neutron stars with large magnetic fields \citep{Dun92}, thus
providing a potential means of coupling the large rotation rate to the
material just outside the neutron star. Calculations so far are
encouraging \citep[e.g.][]{Aki03,Bur07,Jan12b}. No calculation has yet
modeled the full history, of a rotational, or rotational plus neutrino
powered supernova all the way through from the collapse to explosion
phase including all the relevant neutrino and MHD physics, but probably
this will happen in the next decade.

In principle, the outcomes of rotationally powered supernovae and
those powered by neutrinos should be very similar, though only
rotation offers the prospect of making the explosion hyper-energetic
(much greater than 10$^{51}$ erg). To the extent that nucleosynthesis,
light curves and spectra only depend upon the prompt deposition of
$\sim10^{15}$ erg at the center of a highly evolved red or blue
supergiant, they will be indistinguishable. Rotation breaks spherical
symmetry and may produce jets, but except in the case of gamma-ray
bursts, it may be hard to disentangle effects essential to the
explosion from those that simply modify an already successful
explosion. There are interesting constraints on time scales, however,
and hence on field strengths. Rotation or neutrinos must overcome a
ram pressure from accretion that, in the case of high compactness
parameter, may approach a solar mass per second. At a radius of 50 km,
roughly typical of a young hot protoneutron star, it would take a
field strength of over 10$^{15}$ gauss to impede the flow. A similar
estimate comes from nucleosynthesis. In order to synthesize $^{56}$Ni,
material must be heated to at least 4 and preferably $5 \times 10^9$
K. In a hydrodynamical model in which radiation dominates and
10$^{51}$ erg is deposited instantly, this will only occur in a region
smaller than 3000 km.  It takes the shock, moving at typically 20,000
km s$^{-1}$, about 0.1 s to cross that region, after which it begins
to cool off. To deposit 10$^{51}$ erg in that time with a standard
dipole luminosity \citep{Lan80} the field strength would need to exceed
about 10$^{16}$ gauss. This probably exceeds the {\sl surface} fields
generated by collapse alone. Whether the magneto-rotational
instability can generate such fields is unclear, but it may take an
exceptionally high rotation rate for this to all work out.

Perhaps the most common case is a neutrino-powered initial explosion
amplified by rotation at later times. If that is the case though, a
successful outgoing shock must precede any significant pulsar input.
That starting point be difficult to achieve in stars with high
compactness (Figure~\ref{fig:compact}). In any case we do know that
{\sl some} massive stars do make black holes.

\subsection{Magnetar Powered Supernova Light Curves}

\begin{figure}[h]
\centering
\begin{tabular}{cc}
\includegraphics[angle=90,scale=.25]{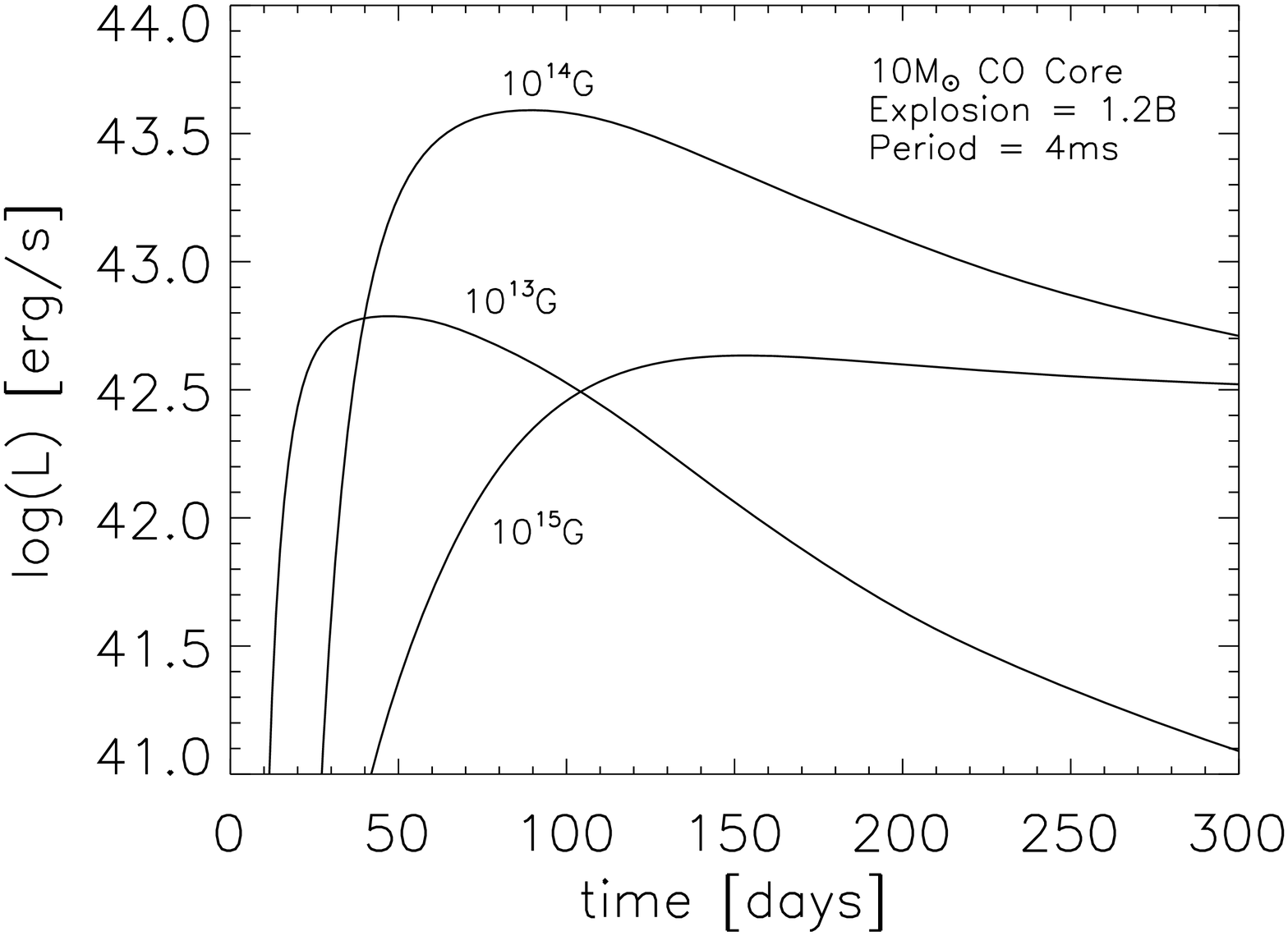} &
\includegraphics[angle=90,scale=.25]{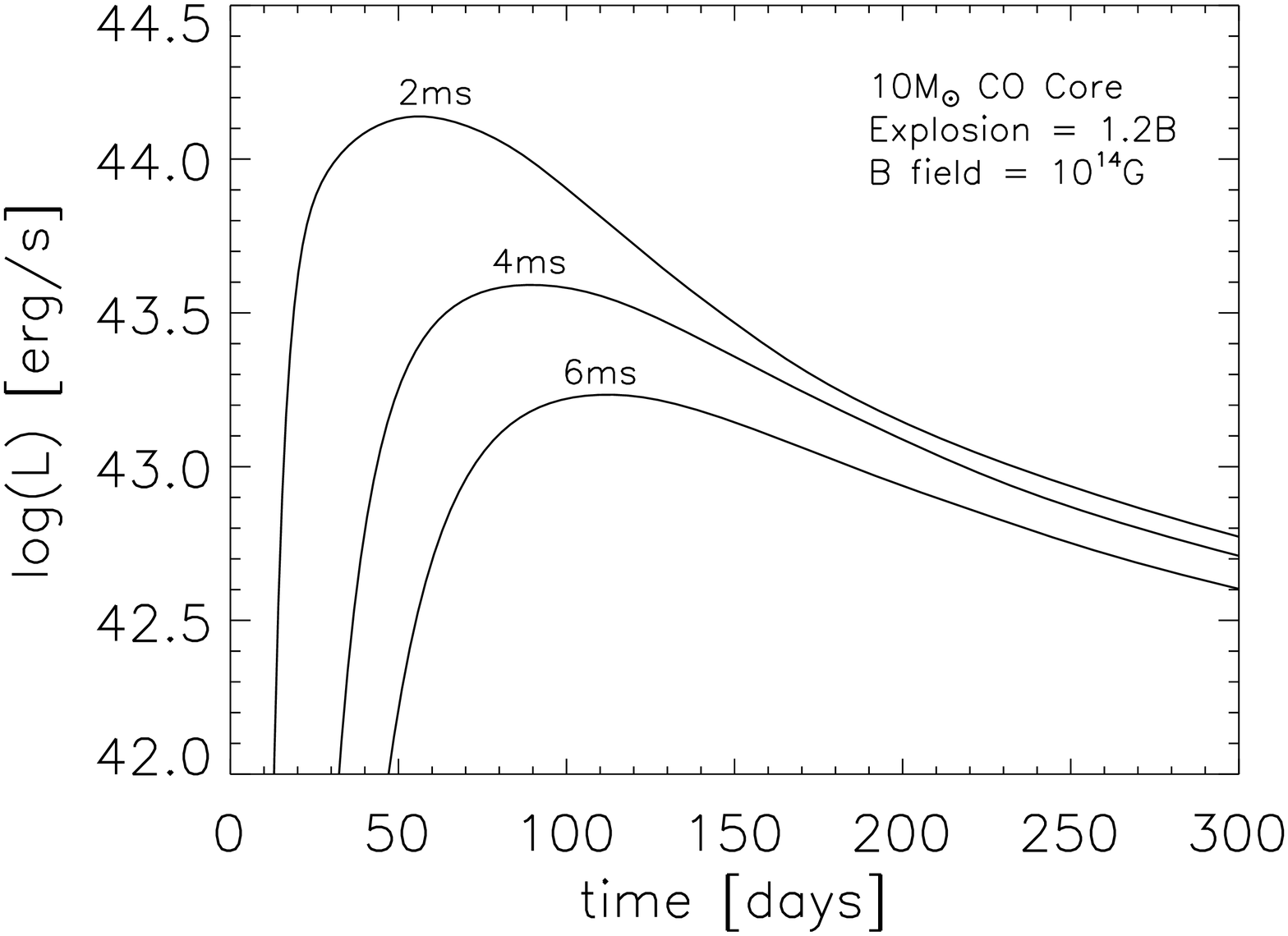} \\
\end{tabular}
\caption{Magnetar powered light curves for (left) different values of
  field strength (10$^{14}$, 10$^{15}$, and 10$^{16}$ G at 4 ms) and
  (right) initial rotation periods (2, 4, 6 ms at 10$^{14}$ G). The
  base event is the 1.2 $\times 10^{51}$ erg explosion of a 10 \Msun
  \ carbon-oxygen core. (Sukhbold and Woosley, 2014, in preparation)}
\label{fig:magnetar}      
\end{figure}

If magnetic fields and rotation can provide the $\sim$10$^{51}$ erg
necessary for the kinetic energy of a supernova, they might, with
greater ease, deliver the 10$^{48}$ or even 10$^{50}$ erg needed to
make a bright - or a really bright - light curve
\citep{Woo10,Kas10}. At the outset, one must acknowledge the huge
uncertainty in applying the very simple pulsar power formula
\citep{Lan80},
\begin{equation}
\frac{dE}{dt} \approx 10^{49} \, B_{15}^2 P_{\rm ms}^{-4} \ {\rm erg \ s^{-1}},
\end{equation}
to a situation where the neutron star is embedded in a dense medium
and that is still be rapidly evolving. Doing this blindly, however,
yields some interesting results (Figure~\ref{fig:magnetar}).  Since
the energy is deposited late, it is less subject to adiabatic losses
and is emitted as optical light with high efficiency. For reasonable
choices of magnetic field and initial rotation rate, the supernova can
be ``ultra-luminous'', brighter than a typical SN Ia for a much longer
time.

The magnetic fields required are not all that large and are similar to
what has been observed for modern day magnetars \citep{Mer08}. In fact,
too large a field results in the rotational energy being deposited too
early. That energy then contributes to the explosion kinetic energy,
but little to the light curve because, by the time the light is
leaking out, the magnetar has already deposited most of its rotational
energy. The rotation rates, though large, are also not extreme, not
very different, in fact, from the predictions for quite massive stars
\citep{Heg05}. If gamma-ray bursts are to be powered by millisecond
magnetars with fields $\sim$10$^{15}$ - 10$^{16}$ G, and if ordinary
pulsars have fields and rotational energies 100 to 1000 times less,
one expects somewhere, sometime to make neutron stars with fields and
rotational energies that are just ten times less. The long
tails on the light curves are interesting and, lacking spectroscopic
evidence or very long duration observations, might easily be confused
with $^{56}$Co decay \citep{Woo10}.

Depending upon the mass and radius of the star, the presence or
absence of a hydrogenic envelope, and the supernova explosion energy,
the resulting magnetar-illuminated transients can be quite
diverse. The brighter events will tend to be of Type I because the
supernova becomes transparent at an earlier time when greater
rotational energy is being dissipated. The upper bound to the
luminosity is a few times 10$^{51}$ erg emitted over several months,
or $\sim$10$^{44.5}$ erg s$^{-1}$, but much fainter events are clearly
possible. For Type II supernovae in red supergiants, the magnetar
contribution may present as a rapid rise in brightness after an
extended plateau \citep{Mae07}. The rise could be even more dramatic
and earlier in a blue supergiant.

An interesting characteristic of 1D models for magnetar powered
supernovae is a large density spike caused by the pile up of matter
accelerated from beneath by radiation. In more than one dimension,
this spike will be unstable and its disruption will lead to additional
mixing that might have consequences for both the spectrum and the
appearance of the supernova remnant.

\subsection{Gamma-Ray Bursts (GRBs)}
\label{sec:grb}

In the extreme case of very rapid rotation and the complete loss of
its hydrogenic envelope, the death of massive star can produce a
common (long-soft) GRB. For a recent review see \citet{Woo13}. There
are two possibilities for the ``central engine'' - a ``millisecond
magnetar'' and a ``collapsar''. The former requires that the product
of a successful supernova explosion be, at least for awhile, a neutron
star, and that the power source is its rotational energy. The latter
assumes the formation of a black hole with a centrifugally supported
accretion disk. The energy source can be either the rotational
energy of that black hole or of the disk, which is, indirectly,
energized by the black hole's strong gravity.

Both models require that the progenitor star have extremely high
angular momentum in and around the iron core. Loss of the hydrogen
envelope could occur though a wind, binary mass exchange, or because
extensive rotationally-induced mixing on the main sequence kept a red
giant from ever forming. Loss of the envelope by a wind is disfavored
because the existence of a lengthy red giant phase would probably
break the rotation of the core to the extent that the necessary
angular momentum was lost. One is this left with the possibility of a
massive star that lost its envelope quite early in to a companion or a
single star that experienced chemically homogeneous evolution
\citep{Mae87,Woo06,Yoo05,Yoo06}. The resulting Wolf-Rayet star must
also not lose much mass or its rotation too will be prohibitively
damped. This seems to exclude most stars of solar metallicity, so GRBs
are relegated to a low metallicity population. The relevant mass
loss rate depends upon metallicity (specifically the iron abundance) as
Z$^{0.86}$ \citep{Vin05}, and even mild reduction is sufficient to
provide the necessary conditions for a millisecond magnetar.

The collapsar model is capable, in principle, of providing much more
energy (up to $\sim$10$^{54}$ erg) than the magnetar model (up to $3
\times 10^{52}$ erg).  The former is limited only by the efficiency of
converting accreted mass into energy, which can be quite high for a
rotating black hole, while the latter is capped by a critical rotation
rate where the protoneutron star deforms and efficiently emits
gravitational radiation. So far, there is no clear evidence for total
(beaming corrected) energies above 10$^{52.5}$ in any GRB, so both
models remain viable. It is interesting that there may be some pile up
of the most energetic GRBs and their associated supernovae around a
few times 10$^{52}$. That might be taken as (mild) evidence in
favor of the magnetar model. On the other hand, black hole production
is likely in the more massive stars and it may be difficult to
arrange things such that all the matter always accretes without
forming a disk \citep{Woo12}

Since angular momentum is in short supply, it is definitely easier to
produce a millisecond magnetar which requires a mass averaged of
angular momentum of only $2 \times 10^{15}$ erg s (for a moment of
inertia I = 10$^{45}$ g cm$^2$), or a value at its equator of $6
\times 10^{15}$ erg s (for a neutron star radius of 10 km).  For
comparison, the angular momentum for the last stable orbit of a Kerr
black hole is $1.5 \times 10^{16} \ \frac{M_{\rm BH}}{3 \ \Msun}$ erg
s and about three times larger for a Schwarzschild hole.  The same
sorts of systems that make collapsars thus also seem likely to make,
at least briefly, neutron stars with millisecond rotation periods. How
these rapid rotators make their fields and how the fields interact
with the rapidly accreting matter in which they are embedded is a very
difficult problem in 3D, general relativistic magnetohydrodynamics.
Analytic arguments suggest however that large fields will be created
\citep{Dun92} and that the rotation and magnetic fields will play a
major role in launching an asymmetric explosion \citep{Aki03,Bur07}.

Just which mass and metallicity stars make GRBs is an interesting
issue. Even when the effects of beaming are included, the GRB event
rate is a very small fraction of the supernova rate and thus the need
for special circumstances is a characteristic of all successful
models.  These special circumstances include, as mentioned, the lack
of any hydrogenic envelope and very rapid rotation.  Without magnetic
torques, the cores of most massive stars would rotate so rapidly at
death that millisecond magnetars, collapsar, and presumably GRBs would
abound. Any realistic model thus includes the effects of magnetic
braking, even though the theory \citep{Spr02,Heg05} is
highly uncertain. In fact, most massive stars may be born with
extremely rapid rotation, corresponding to 50\% critical in the
equatorial plane, because of their magnetic coupling to an accretion
disk \citep{Ros12}. The fact that most massive stars are observed to be
rotating more slowly on the main sequence is a consequence of mass
loss which would be reduced in regions with low metallicity. Since
these large rotation rates are sufficient, again with uncertain
parameters representing the inhibiting effect of composition
gradients, to provoke efficient Eddington-Sweet mixing on the main
sequence, GRBs should be abundant (too abundant?) at low metallicity.
It is noteworthy that models for GRBs that invoke such efficient
mixing on the main sequence do not require that the star be especially
massive since, for low metallicity, the zero age main sequence mass is
not much greater than the presupernova helium core mass
\citep{Woo06}. A low metallicity star of only 15 \Msun \ could become a
GRB and a star of 45 \Msun \ could become a pulsational pair
instability supernova.

Using a standard set of assumptions, the set of massive stars that
might make GRBs by the collapsar mechanism has been surveyed for a
grid of masses and metallicities by \citet{Yoo06}. Averaged over all
redshifts they find a GRB to supernova event ratio of 1/200 which
declines at low redshift to 1/1250. Half of all GRBs are expected to
be beyond redshift 4. Given that magnetars might also make GRBs, or
even most of them, these estimates need to be reexamined.  In
particular, the mean redshift for bursts may be smaller and the
theoretical event rate higher.


\section{Final Comments}

As is frequently noted, we live in interesting times. Most of the
basic ideas invoked for explaining and interpreting massive star death
are now over 40 years old. This includes supernovae powered by
neutrinos, pulsars, the pair-instability, and the pulsational pair
instability. Yet lately, the theoretical models and observational data
have both experienced exponential growth, fueled on the one hand by
the rapid expansion of computer power and the shear number of people
running calculations, and on the other, by large transient
surveys. Ideas that once seemed ``academic'', like pair-instability
supernovae and magnetar-powered supernovae are starting to find
counterparts in ultra-luminous supernovae.

``Predictions'' in such a rapidly evolving landscape quickly become
obsolete or irrelevant. Still, it is worth stating a few areas of
great uncertainty where rapid progress might occur. These issues have been
with us a long time, but problems do eventually get solved.

\begin{itemize}

\item What range(s) of stellar masses and metallicities explode by
  neutrino transport alone. The community has hovered on the brink of
  answering this for a long time. Today some masses explode robustly
  and others show promise \citep{Jan12a,Jan12b}, but a comprehensive,
  parameter-free understanding is still lacking. The computers,
  scientists, and physics may be up to the task in the next five
  years. The compactness of the progenitor very likely plays a major
  role. It would be really nice to know.

\item What is the relation between the initial and final
  (presupernova) masses of stars of all masses and
  metallicities. Suppose we knew the {\sl initial} mass function at
  all metallicities (a big given). What is the {\sl final} mass
  function for presupernova stars?  We can't really answer questions
  about the explosion mechanism of stars of given main sequence masses
  without answering this one too. Our theories and observations of
  mass loss are developing, but still have a long way to go. 


\item What is the angular momentum distribution in presupernova stars?
  To answer this the effects of magnetic torques and mass loss must be
  included throughout all stages of the evolution - a tough
  problem. Approximations exist, but they are controversial and more
  3D modeling might help.

\item Are the ultra-luminous supernovae that are currently being
  discovered predominantly pair instability, pulsational pair
  instability, or magnetar powered (or all three)? Better
  modeling might help, especially with spectroscopic diagnostics.

\item Is the most common form of GRB powered by a rotating neutron
  star or by an accreting black hole? What are the observational
  diagnostics of each?

\item Does ``missing physics'', e.g., neutrino flavor mixing or a
  radically different nuclear equation of state play a role in
  answering any of the above questions?

\end{itemize}

This small list of ``big theory issues'' of course connects to a
greater set of ``smaller issues'' - the treatment of semiconvection,
convective overshoot, and rotational mixing in the models; critical
uncertain nuclear reaction rates; opacities; the complex interplay of
neutrinos, magnetohydrodynamics, convection and general relativity in
3D in a real core collapse - well maybe that is not so small.

Obviously there is plenty for the next generation of stellar
astrophysicists to do.

\section*{Acknowledgements}

We thank Tuguldur Sukhbold and Ken Chen for permission to include
here details of their unpublished work, especially Figures
\ref{fig:compact}, \ref{fig:ppsn2d}, and \ref{fig:magnetar}. This work
has been supported by the National Science Foundation (AST 0909129),
the NASA Theory Program (NNX09AK36G), and the University of California
Lab Fees Research Program (12-LR-237070).

%
%

\begin{thebibliography}{99}%
%
%
\newcommand{\apj}{{\sl ApJ \ }}
\newcommand{\apjl}{{\sl ApJL \ }}
\newcommand{\apjs}{{\sl ApJS \ }}
\newcommand{\physrep}{{\sl Physics Reports\ }}
\newcommand{\baas}{{\sl BAAS\ }}
\newcommand{\baps}{{\sl BAPS\ }}
\newcommand{\nuca}{{\sl Nucl. Phys. A\ }}
\newcommand{\araa}{{\sl Ann. Rev. Astron. Astrophys.\ }}
\newcommand{\aap}{{\sl A\&A \ }}
\newcommand{\prc}{{\sl Phys. Rev. C\ }}
\newcommand{\nat}{{\sl Nature\ }}
\newcommand{\prl}{{\sl PRL\ }}
\newcommand{\apss}{{\sl Astrophys. and Spac. Sci.\ }}
\newcommand{\mnras}{{\sl MNRAS\ }}
\newcommand{\rmp}{{\sl RMP\ }}

\bibitem[Abel et al(2002)]{Abe02} 
Abel, T., Bryan, G.~L., \& Norman, M.~L. \ (2002). Science {\bf 295}, 93. 

\bibitem[Akiyama et al(2003)]{Aki03} 
Akiyama, S., Wheeler, J.~C., Meier, D.~L., \& Lichtenstadt, I.\ (2003).
\apj {\bf 584}, 954. 

\bibitem[Burrows et al(2007)]{Bur07} 
Burrows, A., Dessart, L., Livne, E., Ott, C.~D., \& Murphy,
J.\ (2007). \apj {\bf 664}, 416.

\bibitem[Brown \& Woosley(2013)]{Bro13} 
Brown, J.~M., \& Woosley, S.~E. \ (2013). \apj {\bf 769}, 99.

\bibitem[Chandrasekhar(1939)]{Cha39} 
Chandrasekhar, S. \ (1939). ``An introduction to the study of stellar
structure'',  The University of Chicago press.

\bibitem[Chevalier \& Soderberg(2010)]{Che10} 
Chevalier, R.~A., \& Soderberg, A.~M.\ (2010). \apjl {\bf 711}, L40.

\bibitem[Chieffi \& Limongi(2004)]{Chi04} 
Chieffi, A., \& Limongi, M. \ (2004). \apj {\bf 608}, 405.

\bibitem[Chieffi \& Limongi(2013)]{Chi13} 
Chieffi, A., \& Limongi, M. \ (2013). \apj {\bf 764}, 21.

\bibitem[Chatzopoulos \& Wheeler(2012)]{Cha12} 
Chatzopoulos, E., \& Wheeler, J.~C. \ (2012). \apj {\bf 748}, 42.

\bibitem[Dessart et al(2011)]{Des11} 
Dessart, L., Hillier, D.~J., Livne, E., et al. \ (2011). \mnras, 
{\bf 414}, 2985.

\bibitem[Dessart et al(2012)]{Des12} 
Dessart, L., Hillier, D.~J., Li, C., \& Woosley, S. \ (2012). 
\mnras {\bf 424}, 2139.

\bibitem[Duncan and Thompson(1992)]{Dun92} 
Duncan, R.~C., \& Thompson, C.\ (1992). \apjl {\bf 392}, L9.



\bibitem[Fowler \& Hoyle(1964)]{Fow64} 
Fowler, W.~A., \& Hoyle, F. \ (1964). \apjs {\bf 9}, 201.


\bibitem[Fuller, Woosley, \& Weaver(1986)]{Ful86} 
Fuller, G.~M., Woosley, S.~E., \& Weaver, T.~A.\ (1986). \apj
{\bf 307}, 675.

\bibitem[Galyam et al(2009)]{Gal09} 
Gal-Yam, A., Mazzali, P., Ofek, E.~O., et al., \ (2009). \nat {\bf 462}, 624.

\bibitem[Heger \& Woosley(2002)]{Heg02}
Heger, A., \& Woosley, S.~E., \ (2002). \apj {\bf 567}, 532.

\bibitem[Heger, Woosley, \& Spruit(2005)]{Heg05} 
Heger, A., Woosley, S.~E., \& Spruit, H.~C.\ (2005). \apj {\bf 626}, 350.

\bibitem[Heger \& Woosley(2010)]{Heg10} 
Heger, A., \& Woosley, S.~E.\ (2010). \apj {\bf 724}, 341.

\bibitem[{Hirschi}, {Meynet}, \& {Maeder}(2005)]{Hir05} 
Hirschi, R., Meynet, G., \& Maeder, A. \ (2005). \aap {\bf 433}, 1013.

\bibitem[Hoyle \& Fowler(1960)]{Hoy60} 
Hoyle, F., \& Fowler, W.~A. \ (1960). \apj {\bf 132}, 565.

\bibitem[Janka et al(2012)]{Jan12a}
Janka, H.-T., Hanke, F., H\"udepohl, L., Marek, A., M\"uller, B., \&
Obergaulinger, M. \ (2012). Prog, Theor. Exp. Phys. {\bf 01A309}, 33 pages.

\bibitem[Janka(2012)]{Jan12b}
Janka, H.-T. \ (2012). Ann. Rev. Nucl. and Part. Sci. {\bf 62}, 407.

\bibitem[Kasen \& Woosley(2009)]{Kas09} 
Kasen, D., \& Woosley, S.~E. \ (2009). \apj {\bf 703}, 2205.

\bibitem[Kasen \& Bildsten(2010)]{Kas10} 
Kasen, D., \& Bildsten, L. \ (2010). \apj {\bf 717}, 245.

\bibitem[Lai et al(2008)]{Lai08} 
Lai, D.~K., Bolte, M., Johnson, J.~A., et al. \ (2008). \apj {\bf 681},
1524.

\bibitem[Lang(1980)]{Lan80}
Lang, K. \ (1980). Astrophysical Formulae, (Berlin: Springer).

\bibitem[Lattimer \& Prakash(2007)]{Lat07} 
Lattimer, J.~M., \& Prakash, M.\ (2007). \physrep {\bf 442}, 109.

\bibitem[Limongi, Straniero, \& Chieffi(2000)]{Lim00} 
Limongi, M., Straniero, O., \& Chieffi, A. \ (2000). \apjs {\bf 129}, 625.

\bibitem[Lovegrove \& Woosley(2013)]{Lov13} 
Lovegrove, E., \& Woosley, S.~E.  (2013). \apj {\bf 769}, 109.

\bibitem[Maeda et al(2007)]{Mae07} 
Maeda, K., Tanaka, M., Nomoto, K., et al. \ (2007). \apj {\bf 666}, 1069.

\bibitem[Maeder(1987)]{Mae87} 
Maeder, A. \ (1987).  \aap {\bf 178}, 159.

\bibitem[Maeder \& Meynet(2012)]{Mae12} 
Maeder, A., \& Meynet, G. \ (2012). Reviews of Modern Physics {\bf 84}, 25.

\bibitem[Mereghetti(2008)]{Mer08} 
Mereghetti, S.\ (2008)., Astron and Ap. Rev. {\bf 15}, 225.

\bibitem[Meynet(2002)]{Mey02} 
Meynet, G. \ (2002). \apss {\bf 281}, 183.

\bibitem[M\"uller, Jamka, \& Heger(2012)]{Mul12} 
M{\"u}ller, B., Janka, H.-T., \& Heger, A. \ (2012). \apj {\bf 761} 72.

\bibitem[Nomoto et al(2006)]{Nom06} 
Nomoto, K., Tominaga, N., Umeda, H., Kobayashi, C., \& Maeda, K. \ (2006).
Nuclear Physics A {\bf 777}, 424.

\bibitem[Nomoto, Kobayshi, \& Tominaga(2013)]{Nom13} 
Nomoto, K., Kobayashi, C., \& Tominaga, N. \ (2013). \araa {\bf 51}, 457.

\bibitem[O'Connor \& Ott(2011)]{Oco11} 
O'Connor, E., \& Ott, C.~D. \ (2011). \apj {\bf 730}, 70.

\bibitem[\"Ozel et al(2010)]{Oze10} 
\"Ozel, F., Psaltis, D., Narayan, R., \& McClintock, J.~E.\ (2010). 
\apj {\bf 725}, 1918.

\bibitem[Piro(2013)]{Pir13} 
Piro, A.~L. \ (2013). \apjl {\bf 768}, L14.

\bibitem[Quataert \& Shiode(2012)]{Qua12} 
Quataert, E., \& Shiode, J. \ (2012). \mnras {\bf 423}, L92.

\bibitem[Quataert \& Kasen(2012)]{Qua12b} 
Quataert, E., \& Kasen, D. \ (2012).  \mnras {\bf 419}, L1.

\bibitem[Rosen, Krumholz, \& Ramirez-Ruiz(2012)]{Ros12} 
Rosen, A.~L., Krumholz, M.~R., \& Ramirez-Ruiz, E. \ (2012). \apj {\bf 748}, 
97.


\bibitem[Smartt(2009)]{Sma09} 
Smartt, S.~J.\ (2009). \araa {\bf 47}, 63.

\bibitem[Smartt et al(2009)]{Sma09b} 
Smartt, S.~J., Eldridge, J.~J., Crockett, R.~M., \& Maund, J.~R. \ (2009).
\mnras {\bf 395}, 1409.

\bibitem[Spruit(2002)]{Spr02} 
Spruit, H.~C. \ (2002). \aap {\bf 381}, 923.

\bibitem[Sukhbold \& Woosley(2014)]{Suk14}
Sukhbold, T., \& Woosley, S. E. \ (2014). \apj, in press.


\bibitem[Tan \& McKee(2004)]{Tan04} 
Tan, J.~C., \& McKee, C.~F.\ (2004). \apj {\bf 603}, 383.

\bibitem[Thielemann, Nomoto, \& Hashimoto(1996)]{Thi96} 
Thielemann, F.-K., Nomoto, K., \& Hashimoto, M.-A.\ (1996). \apj 
{\bf 460}, 408.

\bibitem[Timmes, Woosley, \& Weaver(1996)]{Tim96} 
Timmes, F.~X., Woosley, S.~E., \& Weaver, T.~A. \ (1996). \apj 
{\bf 457}, 834.

\bibitem[Ugliano et al(2012)]{Ugl12} 
Ugliano, M., Janka, H.-T., Marek, A., \& Arcones, A. \ (2012).
\apj {\bf 757}, 69.

\bibitem[Vink \& de Koter(2005)]{Vin05} 
Vink, J.~S., \& de Koter, A.\ (2005). \aap {\bf 442}, 587.

\bibitem[Vink et al(2011)]{Vin11} 
Vink, J.~S., Muijres, L.~E., Anthonisse, B., et al.\ (2011).
\aap {\bf 531}, A132.

\bibitem[Vink et al(2013)]{Vin13} 
Vink, J.~S., Heger, A., Krumholz, M.~R., et al. \ (2013).
to be published in Highlights of Astronomy, arXiv:1302.2021.

\bibitem[Wiktorowicz, Belczynski, \& Maccarone(2014)]{Wik14} 
Wiktorowicz, G., Belczynski, K., \& Maccarone, T.~J. \ (2014)
\apj submitted, arXiv:1312.5924.

\bibitem[Woosley(2010)]{Woo10} 
Woosley, S.~E.\ (2010). \apjl {\bf 719}, L204.

\bibitem[Woosley(2013)]{Woo13} 
Woosley, S. E. \ (2013). in Gamma-ray Bursts, by C. Kouveliotou ,
R. A.~M.~J.~Wijers, and S. E. Woosley, (Cambridge University Press,
Cambridge, 191.

\bibitem[Woosley \& Weaver(1995)]{Woo95} 
Woosley, S.~E., \& Weaver, T.~A. \ (1995). \apjs {\bf 101}, 181.

\bibitem[Woosley, Heger, \& Weaver(2002)]{Woo02} 
Woosley, S.~E., Heger, A., \& Weaver, T.~A.\ (2002). Reviews of 
Modern Physics {\bf 74}, 1015.

\bibitem[Woosley \& Heger(2006)]{Woo06} 
Woosley, S.~E., \& Heger, A. \ (2006). \apj {\bf 637}, 914.

\bibitem[Woosley, Blinnikov, \& Heger(2007)]{Woo07}
Woosley, S.~E., Blinnikov, S., \& Heger, A.  \ (2007).
\nat {\bf 450}, 390.

\bibitem[Woosley \& Heger(2007)]{Woo07b} 
Woosley, S.~E., \& Heger, A. \ (2007). \physrep {\bf 442}, 269.

\bibitem[Woosley \& Heger(2012)]{Woo12} 
Woosley, S.~E., \& Heger, A.\ (2012). \apj {\bf 752}, 32.

\bibitem[Yoon \& Langer(2005)]{Yoo05} 
Yoon, S.-C., \& Langer, N. \ (2005). \aap {\bf 443}, 643.

\bibitem[Yoon \& Langer(2006)]{Yoo06} 
Yoon, S.-C., \& Langer, N. \ (2006). \aap {\bf 460}, 199.

\end{thebibliography}
%

%
%
%
%
%

\end{document}